\documentclass[12pt]{article}

\usepackage{authblk}
\usepackage[top=25truemm,bottom=28truemm,left=24truemm,right=24truemm]{geometry}
\usepackage{setspace}
\normalsize
\usepackage{amsmath,amssymb,mathrsfs,amsbsy,latexsym,amsfonts,amsthm,bm,fancyhdr,fancybox,titlesec,ulem,comment,appendix}
\usepackage{physics}
\usepackage{color}
\usepackage{xcolor}

\usepackage[colorlinks=true,linkcolor=blue,citecolor=cyan]{hyperref}

\usepackage{graphicx}
\usepackage{color}
\usepackage{comment}
\usepackage{here}
\usepackage{braket}
\usepackage[utf8]{inputenc}
\usepackage{mathtools}

\newcommand{\nn}{\nonumber\\}

\usepackage{epsf}
\newcommand{\cO}{\mathcal{O}}

\newcommand{\cN}{\mathcal{N}}
\newcommand{\lr}[1]{\left( #1 \right)}
\newcommand{\lrm}[1]{\left\{ #1 \right\}}
\newcommand{\lf}[1]{\lfloor #1 \rfloor}
\newcommand{\ep}{\epsilon}
\numberwithin{equation}{section}
\newcommand{\NT}[1]{{#1}}
\newcommand{\MD}[1]{{#1}}

\newcommand{\NTmod}[1]{{#1}}
\newcommand{\md}[1]{{#1}}
\newcommand{\blue}[1]{{#1}}
\newcommand{\re}[1]{{#1}}

\title{\bf 
Probing Black Hole Thermal Effects \nn in the Dual CFT via Wave Packets}

\author[1]{Norihiro~Tanahashi\thanks{\tt tanahashi(at)gauge.scphys.kyoto-u.ac.jp}}
\author[2]{Seiji~Terashima\thanks{\tt terasima(at)yukawa.kyoto-u.ac.jp}}
\author[1]{Shiki~Yoshikawa\thanks{\tt yoshikawa.shiki(at)gauge.scphys.kyoto-u.ac.jp}}
\vspace{5mm}
\affil[1]{\it\normalsize Department of Physics, Kyoto University, Kyoto 606-8502, Japan}
\affil[2]{\it\normalsize 
Center for Gravitational Physics and Quantum Information,  
\mbox{Yukawa Institute for Theoretical Physics, Kyoto University, Kyoto 606-8502, Japan}  }
\date{\today}

\begin{document}

\maketitle
\thispagestyle{fancy}
\renewcommand{\headrulewidth}{0pt}

\begin{abstract}
We investigate how the gravitational effects of a black hole manifest themselves as thermal behavior in the dual finite-temperature conformal field theory (CFT). In the holographic framework of AdS/CFT, we analyze a wave packet propagating into a black hole geometry in the bulk by computing three-point functions of a scalar primary operator in the boundary CFT. Our setup captures thermal signatures such as exponential damping of the expectation value, which are absent at zero-temperature. This provides a concrete and analytically tractable example of how black hole physics can be probed from the CFT side.
\end{abstract}
\newpage
\thispagestyle{empty}
\setcounter{tocdepth}{2}

\setlength{\abovedisplayskip}{12pt}
\setlength{\belowdisplayskip}{12pt}

\tableofcontents
\newpage

\pagestyle{plain}

\section{Introduction and summary}
Black holes are central objects in the study of quantum gravity, offering a unique window into the interplay between geometry, thermodynamics, and quantum information. In the framework of the AdS/CFT correspondence~\cite{Maldacena:1997re,Gubser:1998bc, Witten:1998qj}, black holes in the bulk spacetime are dual to thermal states in the boundary conformal field theory (CFT), and are therefore expected to encode quantum gravitational effects in a holographically accessible way~\cite{Witten:1998zw, Horowitz:1999jd}. Nevertheless, directly observing the influence of bulk gravitational phenomena on boundary observables remains a significant challenge. While gravitational computations---such as bulk-cone singularities~\cite{Hubeny:2006yu, Maldacena:2015iua}---suggest rich causal and geometric structures in the bulk, their precise CFT counterparts are not fully understood.

A promising approach to this problem is to study wave packets in the bulk spacetime. Wave packets serve as localized, dynamical probes that can test the causal structure and physical response of the geometry. In the case of pure AdS, wave packets have been analyzed extensively~\cite{Terashima:2021klf,Terashima:2023mcr,Tanahashi:2025fqi}, and their dual description in the boundary CFT can be understood via energy densities propagating at the speed of light.\footnote{
Our works and the works in~\cite{Terashima:2021klf, Terashima:2023mcr, Tanahashi:2025fqi} are based on the earlier works on the bulk picture from the CFT side~\cite{Terashima:2020uqu, Terashima:2021klf, Terashima:2017gmc, Terashima:2019wed}, which give some non-traditional claims on the many aspects of the bulk reconstruction in AdS/CFT~\cite{Sugishita:2022ldv, Sugishita:2023wjm, Sugishita:2024lee, Terashima:2025shl}.  
There have been important works on the wave packets in AdS/CFT, for example~\cite{Horowitz:1999gf, Gary:2009ae, Gary:2009mi, Kinoshita:2023hgc, Caron-Huot:2025hmk}, which studied some aspects of them in slightly different ways from ours. }
However, much less is known about the behavior of wave packets in black hole geometries, particularly how their gravitational interactions are encoded in the dual thermal CFT. Although many studies have been conducted from the bulk perspective, it remains difficult to detect direct signatures of black hole effects purely from the CFT side.

In this work, we construct a wave packet that propagates in a three-dimensional bulk geometry containing a BTZ black hole, and analytically compute the corresponding three-point function in the dual two-dimensional finite-temperature CFT. 
Inspired by the method developed in~\cite{Caputa:2014eta}, we evaluate the expectation value of a primary operator for the wave packet state (and the localized excited states). While in two-dimensional CFTs the energy density factorizes into left- and right-moving components and thus shows no thermal decay along the light cone, we find that the expectation value of the primary operator exhibits temperature-dependent exponential decay---a feature absent at zero-temperature.\footnote{
\re{We emphasize that our results do not rely on any large-\(N\) assumption. In particular, the expectation values we obtain hold for a general CFT and thus have broader implications for thermal effects in such systems.}
}

From the bulk point of view, the wave packet is expected to experience a gravitational time delay (Shapiro delay)~\cite{Shapiro:1964uw} due to the presence of the black hole.
Moreover, the thermal effects in the bulk related to the Hawking radiation~\cite{Hawking:1975vcx} should be included.
However, these effects do not manifest in the propagation of energy density on the CFT side, which continues to travel along the light cone without decay. In contrast, we find that thermal effects induced by the black hole can be detected through more sensitive probes, such as the expectation value of primary operators.\footnote{
\re{In other words, the energy-momentum tensor is special in two-dimensional CFTs because it is conserved and chiral, and the corresponding bulk degrees of freedom, namely boundary gravitons, are expected to be localized near the asymptotic boundary. In contrast, for CFTs in three or higher dimensions, even the expectation value of the energy-momentum tensor decays. Expectation values of primary operators, on the other hand, furnish probes that are available in arbitrary dimensions.
}
} 

Our analysis thus provides a concrete and computable example of how gravitational effects in the bulk—specifically thermal damping—can leave observable imprints in boundary correlators, even when more direct signatures like time delay remain elusive.\footnote{
\re{
For the planar black hole, the wave packet cannot reach another boundary point. Even in the BTZ black hole case, this is still impossible because of the special properties of BTZ black holes; in particular, there is an energy gap between the BTZ black hole and the AdS vacuum. By contrast, in higher-dimensional black hole spacetimes that allow bulk trajectories from one boundary point to another, a similar analysis, if feasible, should make it possible to compute in the dual CFT a delay corresponding to the bulk Shapiro delay and bulk-cone singularities.
}
} This framework may serve as a foundation for exploring more complex geometries or for identifying CFT observables that are sensitive to bulk causal structure, potentially deepening our understanding of how gravity, causality, and locality emerge from holographic field theories.

This paper is organized as follows.
In Section \ref{sec:wave-packets}, we construct the wave packet state in the bulk, which serves as the basic probe throughout our analysis.
In Section \ref{sec:3-point}, we compute the expectation values of physical observables
\NTmod{(the expectation value of a scalar primary operator and the energy density)}
in the dual CFT, first in the zero-temperature case corresponding to pure AdS, and then at finite temperature corresponding to a black hole background.
In Section \ref{sec:discussion}, we summarize and discuss the implications of our results \NTmod{in the bulk picture}.
Several technical details and intermediate steps are presented in the Appendix.
\NTmod{Appendix~\ref{shws} analyzes the localized excited state, which can be compared with the wave packet state this work focuses on. Appendices~\ref{gendelt} and \ref{zetagg} show some details of the analysis on the expectation value of the primary operator. Appendix~\ref{ded} summarizes the derivation of the energy density of the wave packet state discussed in~\ref{energy}. Appendix~\ref{unt} is devoted to the discussion on the unitary wave packet operators, which supplements the result in section~\ref{sec:zero-temperature-cft}. }

\section{Wave packets in AdS/CFT}
\label{sec:wave-packets}

We construct wave packet states in the AdS/CFT correspondence following~\cite{Terashima:2023mcr}. The AdS/CFT correspondence implies that 
the bulk Hilbert space in AdS  
corresponds to
the Hilbert space of the dual conformal field theory (CFT). 
Then, we prepare a single-particle state in the bulk theory, which is represented by
a bulk wave packet and is equivalently described as a CFT state.
Specifically, we will focus on 
Gaussian wave packets, which will be used in the following analysis.

In \((d+1)\)-dimensional Minkowski spacetime, the wave packet for a free scalar field $\phi$ at \(t = \Vec{x} = 0\) can be expressed as
\begin{align}
    \int d\Vec{x} \, e^{-\frac{\Vec{x}^2}{2a^2} + i\Vec{p} \cdot \Vec{x}} \phi(t = 0, \Vec{x}) \ket{0},
    \label{eq:minkowski_wavepacket}
\end{align}
where \(\Vec{p}\) represents the momentum of the wave packet.
Instead of specifying the momentum of a space-like direction, say $x_1$, 
we can specify the energy because they are related by the on-shell condition \NT{$\omega^2 - \vec p\,{}^2 = 0$}. 
Then, the wave packet
integrated over time and  spatial directions \(x^i\) (\(i = 2, \ldots, d\)), is given by
\begin{align}
    \int dt \prod_{i = 2}^{d} dx^i \, e^{-\frac{x^i x_i + t^2}{2a^2} + i p_i x^i - i \omega t} \phi(t, \Vec{x})\big|_{x_1 = 0} \ket{0},
    \label{eq:ads_wavepacket}
\end{align}
where \(i\) runs over directions except for 
\(x_1\)
and the size of the wave packet is given by \(a\). Here, we assume \( a \, \sqrt{{\vec p}^2} =  a \,\omega \gg 1\)
to ensure
that the wave packet has a definite 
\NT{propagation direction}
with the momentum \(p_i\) and energy \(\omega\).

Based on the above construction, let us now consider wave packets in the AdS/CFT correspondence,
using
a \MD{scalar} field \(\phi\) in the Poincaré patch of the \((d+1)\)-dimensional AdS spacetime.
In the AdS spacetime, the wave packet can be well-approximated by the Minkowski wave packet if the size of the wave packet \(a\) is much smaller than the AdS radius \(l_{\mathrm{AdS}} = 1\), i.e., \(a \ll 1\). 
At the boundary, the bulk scalar field \(\phi\) is related to the CFT primary operator \(\mathcal{O}\) via the BDHM relation~\cite{Banks:1998dd}.
Thus, we can write the bulk wave packet state at the boundary as  
\begin{align}
    \ket{p, \omega} 
    &= \lim_{z \to 0} \frac{1}{z^\Delta} 
    \int dt \, d^{d-1} x \, e^{-\frac{x^i x_i + t^2}{2a^2} + i p_i x^i - i \omega t} \phi(t, z, x^i) \ket{0} \nonumber \\
    &= \int dt \, d^{d-1} x \, e^{-\frac{x^i x_i + t^2}{2a^2} + i p_i x^i - i \omega t} \mathcal{O}(t, x) \ket{0},
    \label{wads}
\end{align}
where we have identified \(x^1 = z\), and \(z \to 0\) corresponds to the AdS boundary in the Poincaré AdS coordinates. Here, \(\mathcal{O}(t, x)\) is the boundary CFT operator dual to the bulk scalar field \(\phi(t, z, x^i)\).
The wave packet inside the bulk is given by the time evolution of this state \(\ket{p, \omega}\), and 
the time evolution of the bulk wave packet state follows a light-like trajectory. 

The size of the wave packet in spacetime remains \(\mathcal{O}(a)\) during this evolution.
The time evolution of this state in the bulk picture can be evaluated as follows~\cite{Terashima:2017gmc}. 
The bulk localized (one-particle) state is
\begin{align}
\phi(t,z,x^i) \ket{0}=
C \int_{\omega' > \sqrt{k^2}} 
d \omega' d k_i \,  e^{i \omega' t-i k_j x^j } \, z^{\frac{d}{2}} J_{\nu}(\sqrt{ \omega'^2 -k^2 
} \, \, z) 
\, a^\dagger_{\omega', k}  \ket{0},
\label{bs}
\end{align}
where the state is not normalized. 
In order to consider the bulk spatial distribution of the wave packet state \eqref{wads} at time $t$,
we will consider the following overlap:
\begin{align}
& \bra{0} \phi(t=0,z,x^i) \,\,\, e^{i H t} \ket{p,\omega} \nn
= &
 (a \sqrt{\pi})^{d} C^2 \int_{\omega' > \sqrt{k^2}} 
d \omega' d k_i \,  e^{i \omega' t+i k_j x^j } e^{-\frac{a^2}{2} ((k_i-p_i)^2+(\omega'-\omega)^2)} \, 
\sqrt{ \omega'^2 -k^2}^{2 \Delta-d} \, z^{\frac{d}{2}} J_{\nu}(\sqrt{ \omega'^2 -k^2 
} \, \, z),
\label{ov}
\end{align}
where we have used the momentum-space representation of 
\eqref{wads} and the commutation relations of the creation and annihilation operators.
Due to the Gaussian factor, the integrals are dominated by the region near $k_i=p_i, \omega'=\omega$.
Defining $\delta \omega \equiv  \omega' - \omega, \,  \delta k_i \equiv  k_i -p_i$ 
and $p_z \equiv \sqrt{ \omega^2 -p^2 } $,
the overlap can be approximated as
\begin{align}
& \bra{0} \phi(t=0,z,x^i) \,\,\, e^{i H t} \ket{p,\bar{\omega}} \nn
\sim &
 (a \sqrt{\pi})^{d} C^2 \sqrt{2/\pi} z^{\frac{d-1}{2}} (p_z)^{2 \Delta-d-1/2} \,e^{i \omega t+i p_j x^j } \, 
 \nn 
\,\, & \times
 \int 
d \delta \omega d \delta k_i \,  e^{i \delta \omega t+i \delta k_j x^j } e^{-\frac{a^2}{2} ( (\delta k)^2+\delta \omega^2)}  \cos ({(p_z)^2+\omega \delta \omega-p^i \delta k_i \over p_z} z-\frac{2\nu+1}{4} \pi),
\label{ov2}
\end{align}
where we used the asymptotic form of the Bessel function.
The remaining integrals are proportional to 
\begin{align}
 \int 
d \delta \omega d \delta k_i \,  e^{i \delta \omega (t\pm \omega z/p_z)+i \delta k_j (x^j \mp p^j z/p_z) } e^{-\frac{a^2}{2} ( (\delta k)^2+\delta \omega^2)}
\simeq 
e^{-{ 1\over 2 a^2} \left( (t\pm \omega z/p_z)^2+  (x^j \mp p^j z/p_z)^2 \right)},
\label{ov3}
\end{align}
which is strongly suppressed by the Gaussian factor for 
\begin{align}
|t\pm \omega z/p_z| \gg a, \,\,\, \mbox{  or }
\quad
|x^j \mp p^j z/p_z| \gg a.
\end{align}
Therefore, at each time $t>0$, 
the wave packet is localized at $z = { p_z \over \omega} t, \,\, x^j =- {p^j 
 \over \omega} t $, which is on the light-like trajectory from
the boundary point at $t=0$ with the energy $\omega$ and the momentum $p_z, p_i$, as expected. This state has been used to analytically compute the energy density corresponding to a wave packet propagating in the bulk, in the contexts of both AdS\(_3\)/CFT\(_2\) and AdS\(_4\)/CFT\(_3\)~\cite{Terashima:2023mcr,Tanahashi:2025fqi}.

Wave packets have also been used to study the relation between bulk S-matrix elements and CFT correlators in the context of AdS/CFT. In~\cite{Gary:2009ae,Gary:2009mi}, localized bulk wave packets from boundary sources were constructed, and it was investigated how flat-space scattering amplitudes can be extracted from CFT data. 
In contrast to these earlier studies, which are based on the GKPW prescription and analyze correlation functions, the present work is characterized by its use of the operator formalism in AdS/CFT, focusing on the construction and analysis of states. 

We note that this state is not a wave packet state in the CFT,
even for the free CFT, because both the energy and the momenta are fixed, and the on-shell condition is not satisfied generically.
Note also that this state is well-defined because of the smearing in the time direction,
not just the spatial directions~\cite{Nagano:2021tbu}.

\section{Three-point functions}
\label{sec:3-point}

In this section, we compute three-point functions of primary operators in a finite-temperature conformal field theory. These correlators are dual to interaction processes of scalar fields in the bulk, and allow us to probe how thermal effects in the CFT encode gravitational interactions in the presence of a black hole. As a preliminary step and for comparison purposes, we begin by analyzing the simpler case corresponding to pure AdS\(_3\), where the boundary theory is at zero temperature. We then proceed to the finite-temperature case by employing conformal transformations in a two-dimensional CFT. Finally, we also evaluate the energy density associated with the wave packet excitation. 
A similar computation can be performed 
for the localized excited state considered in~\cite{Nozaki:2014hna, Caputa:2014eta}, 
which is briefly summarized in Appendix~\ref{shws}.

\subsection{Zero-temperature CFT}
\label{sec:zero-temperature-cft}

In pure AdS\(_3\), we consider a wave packet injected from the boundary into the bulk and compute the expectation value of the corresponding primary operator in the boundary theory. 
In the AdS\(_3\)/CFT\(_2\) correspondence, the wave packet state can be written as follows:
\begin{align}
     \ket{p, \omega} 
    &= \lim_{z \to 0} \frac{1}{z^\Delta} 
    \int dt \, d x \, e^{-\frac{t^2+x^2}{2a^2}- i \omega t+ip x} \phi(t, x, z) \ket{0} \nonumber \\
    &= \int dt \, d x \, e^{-\frac{t^2+x^2}{2a^2} - i \omega t + ipx} \mathcal{O}(t, x) \ket{0}.
    \label{wads3}
\end{align}
This corresponds to a wave packet of width \(a\), propagating from the origin \NTmod{on the AdS boundary} into the bulk with energy \(\omega\) and momenta \(p\) and \(p_z\). Here, we simply denote the momentum in the \(x\)-direction by \(p\), and we assume the on-shell condition \(\omega^2 - p^2 -p_z^2 = 0\).  In the second line, we apply the BDHM relation to express this state in terms of CFT coordinates and operators. Note that we assume \(a\sqrt{p^2+p_z^2} = a\omega \gg 1\) to ensure that the wave packet has a definite propagation direction with the momentum \(p\) and \(p_z\) and energy \(\omega\). 
Using this state, we can calculate the expectation value of the primary operator \(\cO(t,x)\), which is given by the following expression:
\begin{align}
    &\langle \cO(t,x)\rangle_{0,\mathrm{wp}} 
    \coloneqq\frac{\bra{p,\omega}\cO(t,x)\ket{p, \omega} }{\mathcal{N}^2_0} \nn
    & = \int dt_1 dx_1 e^{-\frac{t_1^2+x_1^2}{2a^2}+i\omega t_1 - ip x_1}\int dt_3 dx_3 e^{-\frac{t_3^2+x_3^2}{2a^2}-i\omega t_3 + ip x_3} \langle0|\cO(t_1,x_1) \cO(t,x) \cO(t_3,x_3) \ket{0} /\mathcal{N}^2_0, \label{zeroo}
\end{align}
where the subscript \(0\) is used to indicate the zero-temperature case.
Here, \(\mathcal{N}_0^2\) is a normalization constant:
\begin{align}
    \mathcal{N}^2_0 &\coloneqq \bra{p,\omega}p,\omega\rangle  \nn
    &=  \int dt_1 dx_1 e^{-\frac{t_1^2+x_1^2}{2a^2}+i\omega t_1 - ip x_1}\int dt_3 dx_3 e^{-\frac{t_3^2+x_3^2}{2a^2}-i\omega t_3 + ip x_3} \langle0|\cO(t_1,x_1)  \cO(t_3,x_3) \ket{0}.
    \label{N}
\end{align}

Below, we will demonstrate the calculation of the numerator in \eqref{zeroo}.
To evaluate the three-point function \(\bra{0}\cO(t_1,x_1)\cO(t,x)\cO(t_3,x_3)\ket{0}\), we analyze it within the Euclidean CFT\(_2\) framework.%
\footnote{
For notational convenience, we occasionally include the subscript \(2\) to indicate the insertion point of the middle operator.
}
In terms of complex coordinates, the three-point function can be written as \\
\(\bra{0}\cO(z_1,\bar{z}_1)\cO(z_2,\bar{z}_2)\cO(z_3,\bar{z}_3)\ket{0}\). We know the exact expression of this:
\begin{align}
    \bra{0}\cO(z_1,\bar{z}_1)\cO(z_2,\bar{z}_2)\cO(z_3,\bar{z}_3)\ket{0} = C_{123}\frac{1}{z_{12}^{\Delta/2} z_{23}^{\Delta/2} z_{31}^{\Delta/2}} \frac{1}{\bar{z}_{12}^{\Delta/2} \bar{z}_{23}^{\Delta/2} \bar{z}_{31}^{\Delta/2}},
\end{align}
where \(z_{ij} =z_i-z_j\) and \(\Delta\) is the conformal dimension of the scalar primary operator \(\cO\). Analytically continuing to Minkowski signature
by replacing
\(z\rightarrow u=t+x\) and \(\bar{z} \rightarrow -v= -(t-x)\),%
\footnote{
This definition of the coordinates $u,v$ is opposite to the standard definition $u=t-x, v = t+x$. We keep using our definition of $u,v$ to facilitate comparison with results of~\cite{Terashima:2023mcr,Tanahashi:2025fqi}.
} 
we obtain the following expression:
\begin{align}
        \bra{0}\cO(u_1,v_1)\cO(u_2,v_2)\cO(u_3,v_3)\ket{0} = C_{123}\frac{1}{u_{12}^{\Delta/2} u_{23}^{\Delta/2} u_{31}^{\Delta/2}} \frac{1}{v_{21}^{\Delta/2} v_{32}^{\Delta/2} v_{13}^{\Delta/2}}.
\end{align}
Here, $C_{123}$ is a constant and is expected to be ${\cal O}(1/N)$ for a holographic CFT, if it is nonzero. 

We also note that only time-ordered correlators make sense in the Euclidean signature. Then,  to reflect this property, we need to use the \(i\epsilon\)-prescription in the analytic continuation to Minkowski signature. In practice, this can be implemented by adding an infinitesimal imaginary-time shift to the operator insertion times. Specifically, the \(i\epsilon\)-prescription is:
\begin{align}
    t_{ij} \rightarrow t_{ij} - i\,\epsilon _{ij}, \qquad \epsilon_{ij} > 0 ~~\mathrm{for}~~ i<j,
\end{align}
as discussed in~\cite{duffin_lecture}.
Then, the expectation value of \(\cO(t,x)\) is computed as
\begin{align}
     & \bra{p,\omega}   \cO(t=t_2, x=x_2) \ket{p,\omega} \nn
   =
   & \int d t_1 \, d x_1 \, 
   e^{-\frac{ t_1^2+x_1^2}{2 a^2}-i p x_1+i {\omega} t_1} 
   \int d t_3 \, d x_3 \, 
   e^{-\frac{ t_3^2 + x_3^2}{2 a^2}+i p x_3-i {\omega} t_3}
   \nn
   &\times C_{123}\frac{1}{u_{12}^{\Delta/2} u_{23}^{\Delta/2} u_{31}^{\Delta/2}} \frac{1}{v_{21}^{\Delta/2} v_{32}^{\Delta/2} v_{13}^{\Delta/2}} \nn
   =&  \frac{1}{4}\int d u_1 \, d v_1 \, d u_3 \, d v_3 \,\, e^{-\frac{ (u_1-i a^2p_u)^2+(v_1-i a^2p_v)^2+(u_3+i  a^2p_u)^2+(v_3+i a^2p_v)^2}{4 a^2} 
    - \frac{a^2}{2} ((p_u)^2+(p_v)^2) } \nn
    & \times C_{123}\frac{1}{u_{12}^{\Delta/2} u_{23}^{\Delta/2} u_{31}^{\Delta/2}} \frac{1}{v_{21}^{\Delta/2} v_{32}^{\Delta/2} v_{13}^{\Delta/2}},
\end{align}
where we defined \(p_u \coloneqq \omega-p \,\, (\gg0)\) and \(p_v \coloneqq \omega + p \,\,(\gg 0)\).

\begin{figure}
    \centering
    \includegraphics[width=0.5\linewidth]{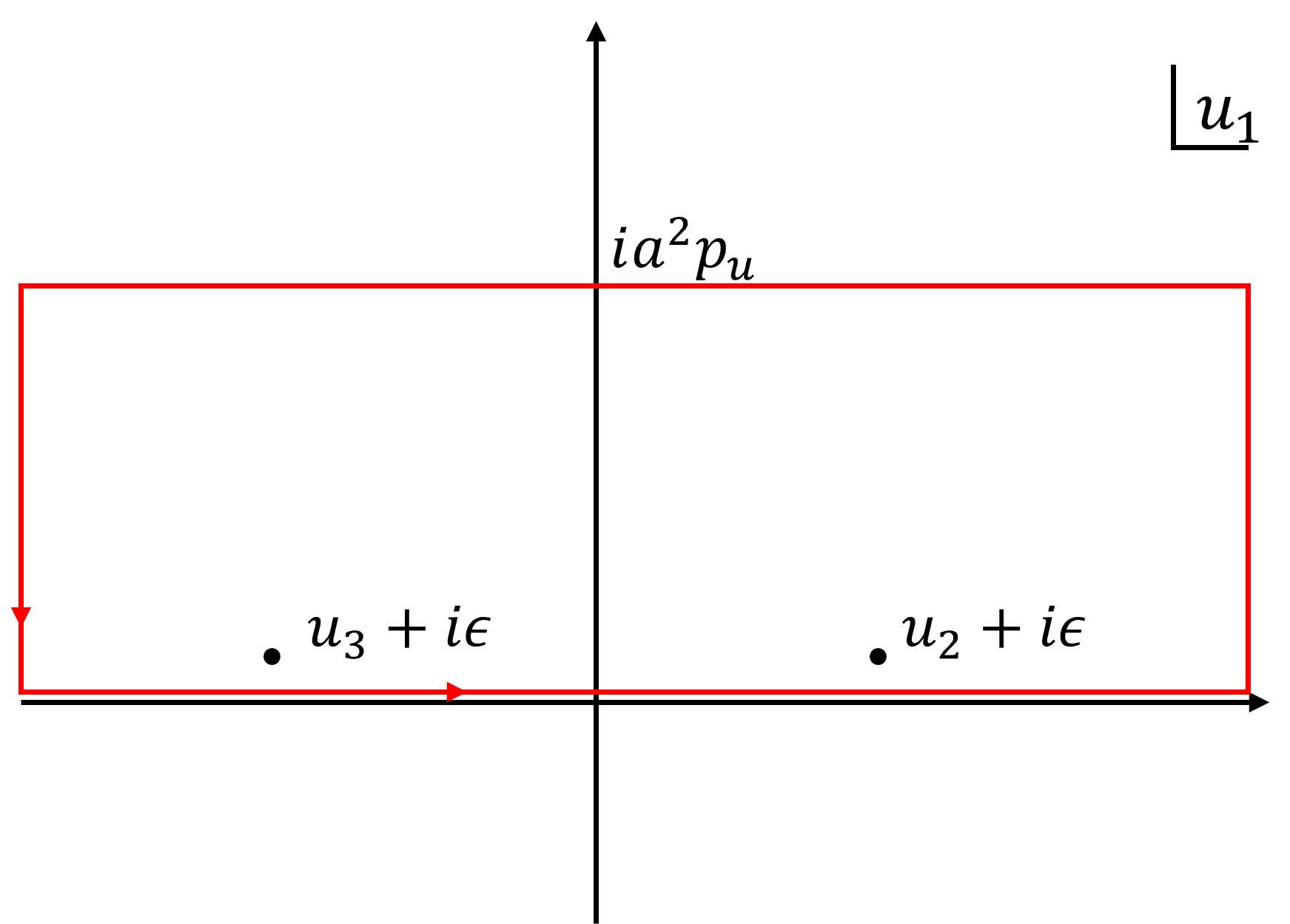}
    \caption{\(u_1\)-integration contour.}
    \label{f1}
\end{figure}

We evaluate the \(u\)-part of the expression. 
\MD{Although we can evaluate it for arbitrary conformal dimension \(\Delta\), we focus on \(\Delta=2\) for simplicity. In appendix~\ref{gendelt}, we evaluate it for generic~\(\Delta\).}%
\footnote{
In~\cite{Terashima:2023mcr,Tanahashi:2025fqi}, the leading momentum-dependent contribution of energy densities was calculated for general \(\Delta\). Similarly to the energy density computation, the leading-order contribution to the expectation value of the primary operator can be evaluated by the same procedure. While the case \(\Delta \neq 2\) leads to a more complicated calculation, the same method applies. For non-integer \(\Delta\), the branch cut structure requires a suitable contour deformation to evaluate the integral as done in~\cite{Terashima:2023mcr}.
}
First, we will compute
\begin{align}
    A \coloneqq& \int du_1 du_3\, e^{-\frac{ (u_1-i a^2p_u)^2+(u_3+i a^2p_u)^2}{4 a^2} 
    - \frac{a^2p_u^2}{2} }
    \frac{1}{\lr{u_1-u_2-i\ep} \lr{u_2-u_3-i\ep} \lr{u_3-u_1+i\ep}} \nn
    \simeq& \int du_3 \lr{ -(2\pi i)\frac{1}{\lr{u_3-u_2+i\ep}^2}e^{-\frac{\lr{u_2-ia^2p_u}^2}{4a^2}}e^{-\frac{\lr{u_3+ia^2p_u}^2}{4a^2}}e^{-\frac{a^2p_u^2}{2}} 
    +(2\pi i)\frac{e^{-\frac{u_3^2}{2a^2}}}{\lr{u_3-u_2+i\ep}^2}} \nn
    \simeq & \, 2\pi^2 e^{-\frac{u_2^2}{2a^2}} 
    \NTmod{(ip_u + u_2/a^2)}
    +2\pi i \frac{\sqrt{2\pi}a}{\lr{u_2}^2}
    \Bigl(1+{\cal O}\bigl((a/u_2)^2\bigr)\Bigr)
    \NTmod{- 2\pi^2 e^{-\frac{u_2^2}{2a^2}}\frac{u_2}{a^2}}
    ,
    \label{A}
\end{align}
where the first term was obtained by the residue theorem, and the second term was obtained by the saddle-point approximation. 
The last term in the third line, which is proportional to $u_2$, originates from
the pole contribution.
This term is real and canceled by the corresponding part of the first term.
Note that we neglect ${\cal O}(e^{-(a p_u)^2/4})$ terms since $a\, p_u \gg 1$.
Below, 
we will take  \(|u|\gg a\) and
omit the terms of ${\cal O}((a/u)^2)$ in \eqref{A} for notational simplicity.
We will also neglect ${\cal O}(u/(p_u a^2))$ terms by assuming $|u|/(p_u a^2) \ll 1$.\footnote{
As explained above, the result here is valid without this assumption. For the $\Delta \neq 2$ case and the finite temperature case, the results may be valid without this assumption, although we have not checked them.
\NTmod{If this assumption is introduced, the domain in which the analysis here is also bounded from above as $1 \ll u/a \ll a\,p_u$.}
}
An example of the integration contour is shown in Figure~\ref{f1}.%
\footnote{
Thanks to \(i\ep\)-prescription, the pole is located inside the contour. Since the contribution from the upper contour is exponentially suppressed as \(e^{-\frac{a^2p_u^2}{2}}\), it can safely be neglected.
\NTmod{The integration contour for the $u_3$ integration is similar to that in Figure~\ref{f1}, while it is taken in the lower-half complex plane in the clockwise direction.}
}

\begin{figure}
    \centering
    \includegraphics[width=0.7\linewidth]{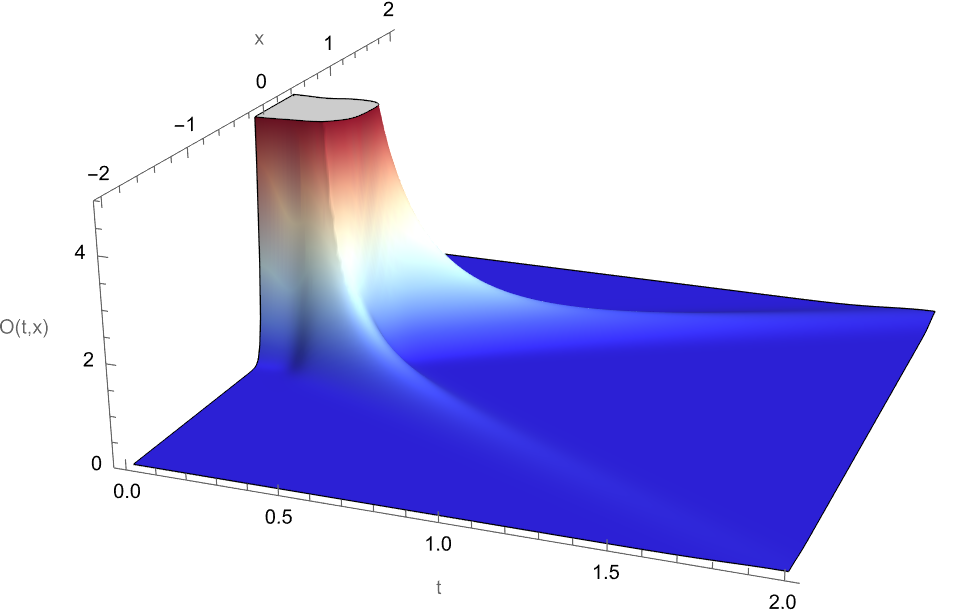}
    \caption{
    \NTmod{Profile of $\langle \cO(t,x)\rangle_{0,\mathrm{wp}}$ given by \eqref{expO0} for $a=0.15, p_u =10, p_v = 25, \ep= 0.1$.}
    \md{The contribution from the fourth term, which spreads spatially, is neglected.}
    The shockwave-induced contributions decay as a power law along the light cone at zero-temperature.}
    \label{f2}
\end{figure}

Apart from the overall sign, the integration over \(v\) proceeds in exactly the same way, and the result is summarized as
\begin{align}
     \bra{p,\omega}   \cO(t=t_2, x=x_2) \ket{p,\omega} 
     \simeq 
     C_{123}\lr{{i\pi^2}p_u e^{-\frac{u_2^2}{2a^2}}+\pi i \frac{\sqrt{2\pi}a}{\lr{u_2}^2}}
     \lr{-i\pi^2p_v e^{-\frac{v_2^2}{2a^2}}-\pi i \frac{\sqrt{2\pi}a}{\lr{v_2}^2}}.
\end{align}
Normalization constant \(\cN_0^2\) can also be calculated as
\begin{align}
    \mathcal{N}^2_0 \coloneqq& \bra{p,\omega}p,\omega\rangle  \nn
    =&  \frac{1}{4}\int d u_1 \, d v_1 \, d u_3 \, d v_3 \,\, e^{-\frac{ (u_1-i a^2p_u)^2+(v_1-i a^2p_v)^2+(u_3+i  a^2p_u)^2+(v_3+i a^2p_v)^2}{4 a^2} 
    - \frac{a^2}{2} ((p_u)^2+(p_v)^2) }\nn
    & \times\frac{1}{\lr{u_1-u_3-i\ep}^2\lr{v_3-v_1+i\ep}^2}\nn
     \simeq & \frac{\pi^3a^2 p_up_v}{2}.
\end{align}
Then, we obtain the expectation value of the primary operator \(\cO(t,x)\):
\begin{align}
    \langle \cO(t,x)\rangle_{0,\mathrm{wp}}
    \simeq
    2C_{123}\lr{\frac{\pi}{a^2}e^{-\frac{u^2}{2a^2}}e^{-\frac{v^2}{2a^2}}
    +\frac{\sqrt{2\pi}}{ap_v v^2}e^{-\frac{u^2}{2a^2}}
    +\frac{\sqrt{2\pi}}{ap_u u^2}e^{-\frac{v^2}{2a^2}}
    +\frac{2}{p_up_v u^2 v^2}}.
    \label{expO0}
\end{align}
The first term in \eqref{expO0} may represent contributions from the excitation due to the wave packet prepared near the origin.  
This is ${\cal O}(C_{123}/a^2)$ and
localized near the origin $x=0$, at $t=0$, but almost vanishes for $t \gg |a|.$
The second and third terms are localized on the light cone and decay as a power law along the light cone. In the zero-temperature case, the second and third terms are suppressed by a factor of \( 1/(a p) \) compared to the first term in the overlapped region near $|u| \sim a, |v| \sim a$. 
This behavior can be observed 
in Figure~\ref{f2}.
The fourth term is proportional to $1/(t^2-x^2)^2$ and large near the light cone, but not localized.
It becomes nonzero in regions that are spacelike separated from the origin. 
This does not contradict causality, because a local operator acting on a state can change the expectation value everywhere.

At $t=0$, \eqref{expO0} has the narrow Gaussian peaks at $x=0$ and smaller $1/x^4$ distributions.
By the time evolution, they decay as a power law, as stated above.
The decay may be caused by the spread of the localized excitation or by diffusion, which occurs even in a vacuum.
The initial state is rather special. The energy and the momentum are almost fixed, but do not satisfy a usual on-shell condition from the CFT viewpoint.
Because it is not a generic localized excitation, the diffusion is not of a generic form, and the light cone structure appears.

In contrast to the one-particle wave packet used here, if we consider the state formed by applying a unitary operator (obtained by exponentiating the anti-Hermitian wave-packet operator) to the vacuum, the contributions corresponding to the fourth term vanish as required by causality.
The details are shown in Appendix~\ref{unt}.

\subsection{Finite-temperature CFT}\label{ftcft}
Having completed the analysis of the zero-temperature case, we consider a bulk spacetime that contains a black hole.
\footnote{
\re{
For technical reasons, we restrict ourselves to the planar black hole geometry in the present paper. In this setup, the black-hole saddle is the relevant dominant one, and thermal AdS does not affect the analysis.
}
}
Our aim is to understand the gravitational effects experienced by wave packets propagating through such a geometry from the perspective of the boundary CFT. To this end, we study the corresponding finite-temperature CFT, where thermal effects are expected to emerge as a manifestation of the bulk black hole. In two-dimensional CFTs, the finite-temperature theory can be obtained from the zero-temperature one by a conformal transformation.\footnote{
In~\cite{Caputa:2014eta}, the energy density in the case of a local quench was analyzed, although thermal effects may not be extracted from the energy density in two-dimensional CFTs,
as 
shown in 
section~\ref{energy}. 
In~\cite{Caputa:2014eta}, the thermal effects were obtained through the holographic computation of entanglement entropy in the bulk. Instead of the entanglement entropy, we will consider the expectation value of the scalar.}

Our goal here is to compute the expectation value of a primary operator in a finite-temperature CFT. To achieve this, we proceed as follows. First, we obtain the expression for the correlator \(\langle\cO(w_1,\bar{w_1})\cO(w_2,\bar{w}_2)\cO(w_3,\bar{w}_3)\rangle_\beta\) in a thermal state by applying a conformal transformation to the corresponding zero-temperature correlator. Next, we analytically continue to Minkowski spacetime, including the appropriate \(i\ep\)-prescription. Finally, we perform the integration with a wave packet smearing.
\subsubsection*{From zero to finite temperature}
To obtain correlation functions at finite temperature, it suffices to consider a conformal transformation from the plane to the cylinder.%
\footnote{
This method strongly relies on the fact that we are working with a two-dimensional CFT.
}
We first introduce the coordinates on the Euclidean plane as follows:
\begin{align}
    z = x'+i\tau', \,\,\,
    \bar{z} = x' - i\tau' .
\end{align}
Since we are interested in finite-temperature physics, we apply a conformal transformation from the plane to the cylinder:
\begin{align}
        z= \exp\left(\frac{2\pi w}{\beta}\right), \;\;\;
    \bar{z} = \exp \left(\frac{2\pi \bar{w}}{\beta}\right),
\end{align}
where \(\beta\) is the inverse temperature, and the Euclidean time \(\tau\) is periodic with period \(\beta\). The complex coordinates \(w\) and \(\bar{w}\) are defined in terms of Euclidean time \(\tau\) as
\begin{align}
        w = x + i\tau , \;\;
    \bar{w} = x - i \tau.
\end{align}
The primary operator \(\cO\) transforms as follows:
\begin{align}
        \cO(w,\bar{w})  = \frac{d z}{dw}^h \frac{d\bar{z}}{d\bar{w}}^{\bar{h}} \cO(z,\bar{z}) 
     = \left(\frac{2\pi}{\beta}\right)^\Delta z^\frac{\Delta}{2} \bar{z}^\frac{\Delta}{2} \cO(z,\bar{z}),
\end{align}
where \(\Delta = h + \bar{h}\). Using these, we obtain the following expression:
\begin{equation}
         \langle \cO(w_1,\bar{w}_1)\cO(w_2,\bar{w}_2)\rangle_\beta  
        =\left( \frac{\pi}{\beta}\right)^{2\Delta} 
            \frac{1}{ \sinh^\Delta \frac{\pi(w_1-w_2)}{\beta}\sinh^\Delta \frac{\pi(\bar{w}_1-\bar{w}_2)}{\beta}}, 
\end{equation}
\begin{align}
    &\langle \mathcal{O}_1(w_1, \bar{w}_1) \mathcal{O}_2(w_2, \bar{w}_2) \mathcal{O}_3(w_3, \bar{w}_3) \rangle_\beta\nn
    &= \, C_{123}\lr{\frac{\pi}{\beta}}^{3\Delta} \frac{1}{ \sinh^{\frac{\Delta}{2}} \left( \frac{\pi (w_1 - w_2)}{\beta} \right)} \frac{1}{ \sinh^{\frac{\Delta}{2}} \left( \frac{\pi (w_2 - w_3)}{\beta} \right)} \frac{1}{ \sinh^{\frac{\Delta}{2}} \left( \frac{\pi (w_3 - w_1)}{\beta} \right)} \nn
    &\qquad\qquad\quad~~ \times
    \frac{1}{ \sinh^{\frac{\Delta}{2}} \left( \frac{\pi (\bar{w}_1 - \bar{w}_2)}{\beta} \right)} 
    \frac{1}{ \sinh^{\frac{\Delta}{2}} \left( \frac{\pi (\bar{w}_2 - \bar{w}_3)}{\beta} \right)}\frac{1}{ \sinh^{\frac{\Delta}{2}} \left( \frac{\pi (\bar{w}_3 - \bar{w}_1)}{\beta} \right)}.
\end{align}
Here, we used the following relations:
\begin{align}
    z_i-z_j = & e^\frac{2\pi w_i}{\beta}-e^\frac{2\pi w_j}{\beta}
    =  2e^\frac{\pi(w_i+w_j)}{\beta} \sinh \frac{\pi(w_i-w_j)}{\beta}.
\end{align}
\subsubsection*{Analytic continuation and \(i\ep\)-prescription}
 \MD{Analytically} continuing to Minkowski signature, we will replace \(z\rightarrow u=t+x\) and \(\bar{z} \rightarrow -v= -(t-x)\) with \(i\ep\)-prescription. We obtain the following expression:
\begin{align}
 &\langle \cO(u_1,v_1)\cO(u,v)\cO(u_3,v_3)\rangle{_\beta}  \nn
    =& C_{123}\lr{\frac{\pi}{\beta}}^{3\Delta} \frac{1}{ \sinh^{\frac{\Delta}{2}} \left( \frac{\pi (u_1 - u-i\ep)}{\beta} \right)} \frac{1}{ \sinh^{\frac{\Delta}{2}} \left( \frac{\pi (u - u_3-i\ep)}{\beta} \right)} \frac{1}{ \sinh^{\frac{\Delta}{2}} \left( \frac{\pi (u_3 - u_1+2 i\ep)}{\beta} \right)} \nn
        &\quad \times
        \frac{1}{ \sinh^{\frac{\Delta}{2}} \left( \frac{\pi (v - v_1+i\ep)}{\beta} \right)} 
        \frac{1}{ \sinh^{\frac{\Delta}{2}} \left( \frac{\pi (v_3 - v+i\ep)}{\beta} \right)}\frac{1}{ \sinh^{\frac{\Delta}{2}} \left( \frac{\pi (v_1 - v_3-2 i\ep)}{\beta} \right)}.
\end{align}

\subsubsection*{Integration}
\begin{figure}
    \centering
    \includegraphics[width=0.5\linewidth]{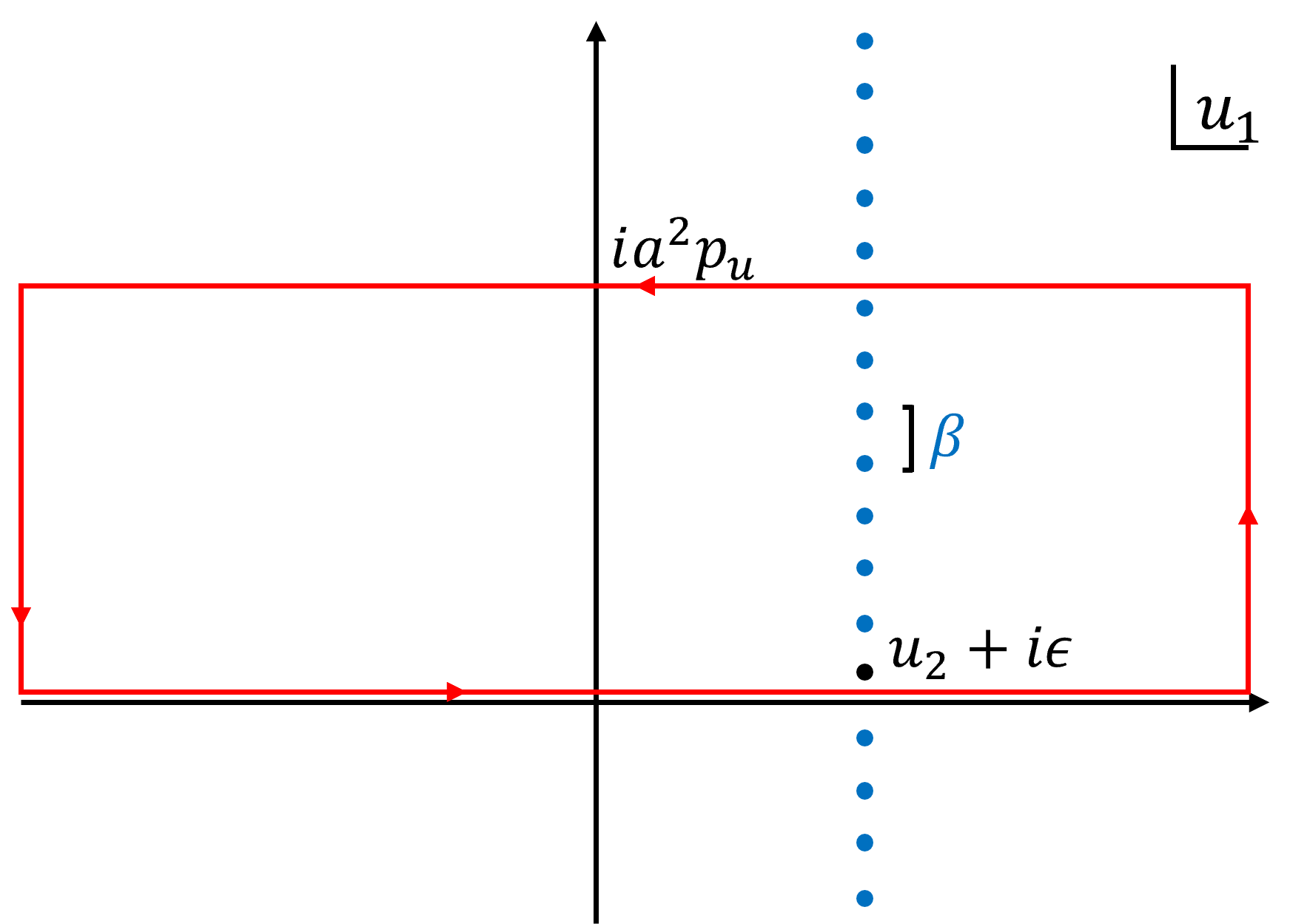}
    \caption{The integration contour and the locations of the poles at finite temperature. The blue dots represent the additional poles that emerge at finite temperature, forming an infinite series along the imaginary axis with a separation of \(\beta\).}
    \label{f8}
\end{figure}

We consider the smeared integral with a wave packet. Here, we focus on the case \(\Delta=2\) for simplicity, although the results will be qualitatively the same for other $\Delta$. We compute the following quantity:
\begin{align}
    &\frac{1}{4}\int du_1 dv_1 du_3 dv_3 e^{-\frac{(u_1-ip_u a^2)^2 + (v_1-ip_va^2)^2+(u_3+ip_ua^2)^2+(v_3+ip_va^2)^2}{4a^2}
    -\frac{a^2p_u^2}{2}-\frac{a^2p_v^2}{2}} \nn
        &\quad \times  C_{123}\lr{\frac{\pi}{\beta}}^{6} \frac{1}{ \sinh \left( \frac{\pi (u_1 - u -i\ep)}{\beta} \right)} \frac{1}{ \sinh \left( \frac{\pi (u - u_3-i\ep)}{\beta} \right)} \frac{1}{ \sinh \left( \frac{\pi (u_3 - u_1+2 i\ep)}{\beta} \right)} \nn
        &\quad \times
        \frac{1}{ \sinh \left( \frac{\pi (v - v_1+i\ep)}{\beta} \right)} 
        \frac{1}{ \sinh \left( \frac{\pi (v_3 - v+i\ep)}{\beta} \right)}
        \frac{1}{ \sinh \left( \frac{\pi (v_1 - v_3-2 i\ep)}{\beta} \right)}. 
\end{align}
At first, we will focus on \(u_1\)-integral.
\begin{align}
    F_1\coloneqq&\int du_1 \, e^{-\frac{(u_1-ip_u a^2)^2}{4a^2}}
        \frac{1}{ \sinh^{\frac{\Delta}{2}} \left( \frac{\pi (u_1 - u -i\ep)}{\beta} \right)} \frac{1}{ \sinh^{\frac{\Delta}{2}} \left( \frac{\pi (u_3 - u_1 +2 i\ep)}{\beta} \right)} \nn
     \simeq & (2\pi i) \sum_{n=0}^{\lf{\zeta}}\frac{\beta}{\pi}
        \frac{e^{-\frac{(u+i\beta(n-\zeta))^2}{4a^2}}}{\sinh \left( \frac{\pi (u_3 - u +i\ep)}{\beta} \right)} 
        + (2\pi i) \sum_{m=0}^{\lf{\zeta}}
       \lr{-\frac{\beta}{\pi}}\frac{e^{-\frac{(u_3+i\beta(m-\zeta))^2}{4a^2}}}{\sinh \left( \frac{\pi (u_3 - u+i\ep)}{\beta} \right)},
       \label{F1}
\end{align}
where 
\begin{align}
    \zeta = \frac{a^2p_u}{\beta}.
\end{align}
\MD{
This sum on \(n\) arises from the additional poles that appear at finite temperature. The integration contour and the locations of the poles are shown in Figure~\ref{f8}.
}

We will consider the following three cases separately: $\zeta \ll 1$,  $\zeta \sim {\cal O}(1)$ and $\zeta \gg 1$.
With the assumption $ a |p| \gg 1$, the cases $\zeta \sim {\cal O}(1)$ and $\zeta \ll 1$ are interpreted as low temperature cases: $a/\beta \ll 1$
(we assume $p_u$ and $p_v$ are of the same order).

First, let us consider the $\zeta \ll 1$ case. 
For this case, only the pole nearest to the real axis contributes, and we have 
\begin{equation}
    F_1
     \simeq  2\pi i
     \NTmod{ \, \frac{\beta}{\pi} \, }
        \frac{e^{-\frac{(u-i\beta \zeta)^2}{4a^2}}}{\sinh \left( \frac{\pi (u_3 - u +i\ep)}{\beta} \right)} 
        + 2\pi i
       \lr{-\frac{\beta}{\pi}}\frac{e^{-\frac{(u_3-i\beta \zeta)^2}{4a^2}}}{\sinh \left( \frac{\pi (u_3 - u+i\ep)}{\beta} \right)}.
\end{equation}
Next, we consider the \(u_3\)-integration.
\begin{align}
    F_3 &\coloneqq \int du_3 \, F_1 e^{-\frac{(u_3+ip_ua^2)^2}{4a^2}
        -\frac{a^2p_u^2}{2}} \frac{1}{\sinh \frac{\pi(u-u_3- i \epsilon)}{\beta}} \nn
    & = \int du_3 \, e^{-\frac{(u_3+ip_ua^2)^2}{4a^2}
        -\frac{a^2p_u^2}{2}}(-1) \nn
        &\quad
        \times \Bigg\{  2\pi i
        \NTmod{ \, \frac{\beta}{\pi} \, }
        \frac{e^{-\frac{(u-i\beta \zeta)^2}{4a^2}}}{\sinh^2 \left( \frac{\pi (u_3 - u +i\ep)}{\beta} \right)} 
        + (2\pi i) 
       \lr{-\frac{\beta}{\pi}}\frac{e^{-\frac{(u_3-i\beta \zeta)^2}{4a^2}}}{\sinh^2 \left( \frac{\pi (u_3 - u+i\ep)}{\beta} \right)} \Bigg\} \nn
    & 
    \simeq 4 \pi^2 
        \lr{\frac{\beta}{\pi}}^3
        \, \frac{1}{2a^2}i\beta \zeta \,
        e^{-\frac{1}{2a^2} u^2} 
        + 2\pi i \, \frac{\beta}{\pi} \, \sqrt{2\pi} a \,
        \frac{1}{
        \sinh^2 
        \lr{\frac{\pi u}{\beta}} 
        }. 
        \label{F3a}
\end{align}
The first term was obtained by the residue theorem, while the second term was obtained via the saddle-point approximation. Note that we neglect terms suppressed by \( a^2/u^2\)
\NTmod{and $u/\beta\zeta$}. 
Apart from the overall sign, the integration over \(v\) proceeds in exactly the same way, and the result is summarized as follows:
\begin{align}
    &\langle\cO(t,x) \rangle_{\beta,\mathrm{wp}}\nn
    = & 4C_{123}\pi^4\lrm{e^{-\frac{u^2}{2a^2}}\frac{ip_u}{2}
    +\frac{i\pi  a}{\beta^2}\sqrt{\frac{\pi}{2}}\lr{\frac{1}{\sinh^2\frac{\pi u}{\beta}} }}
    \times\lrm{-e^{-\frac{v^2}{2a^2}}\frac{ip_v}{2}
    -\frac{i\pi  a}{\beta^2}\sqrt{\frac{\pi}{2}}\lr{\frac{1}{\sinh^2\frac{\pi v}{\beta}} }}/\mathcal{N}_\beta^2, \nn
    \label{obetaa}
\end{align}
where \(\cN_\beta^2\) is defined as
\begin{align}
        \mathcal{N}^2_\beta &\coloneqq \frac{1}{4}\int du_1 dv_1 du_3 dv_3 e^{-\frac{(u_1-ip_u a^2)^2 + (v_1-ip_va^2)^2+(u_3+ip_ua^2)^2+(v_3+ip_va^2)^2}{4a^2}
    -\frac{a^2p_u^2}{2}-\frac{a^2p_v^2}{2}} \nn
    & \quad
    \times \left( \frac{\pi}{\beta}\right)^{2\Delta} 
            \frac{1}{ \sinh^\Delta \frac{\pi(u_1-u_3-i\epsilon)}{\beta}\sinh^\Delta \frac{\pi(v_3 -v_1 + i\epsilon)}{\beta}} \nn
    &\simeq \frac{\pi^3 a^2}{2^{2\Delta-3}}\frac{(p_u p_v)^{\Delta-1}}{\Gamma(\Delta)^2}
     =  \frac{\pi^3 a^2}{2} \, p_u p_v \,.
     \label{OfiniteT}
\end{align}
Then, we arrive at the following expression:
\begin{align}
    \langle\cO(t,x) \rangle_{\beta,\mathrm{wp}}
    &\simeq 2C_{123} \Bigg(
    e^{-\frac{u^2}{2a^2}}e^{-\frac{v^2}{2a^2}}\frac{\pi}{a^2}
    + e^{-\frac{u^2}{2a^2}}\frac{\pi^2}{ap_v\beta^2}\sqrt{2\pi}\frac{1}{\sinh^2\frac{\pi v}{\beta}}\nn
    & \hphantom{\simeq 2C_{123}}
    + e^{-\frac{v^2}{2a^2}}\frac{\pi^2}{ap_u\beta^2}\sqrt{2\pi}\frac{1}{\sinh^2\frac{\pi u}{\beta}}
    +\frac{2\pi^4}{\beta^4 p_up_v}\frac{1}{\sinh^2\frac{\pi u}{\beta}\sinh^2\frac{\pi v}{\beta}}
    \Bigg).
    \label{evo}
\end{align}
\begin{figure}
    \centering
    \includegraphics[width=0.5\linewidth]{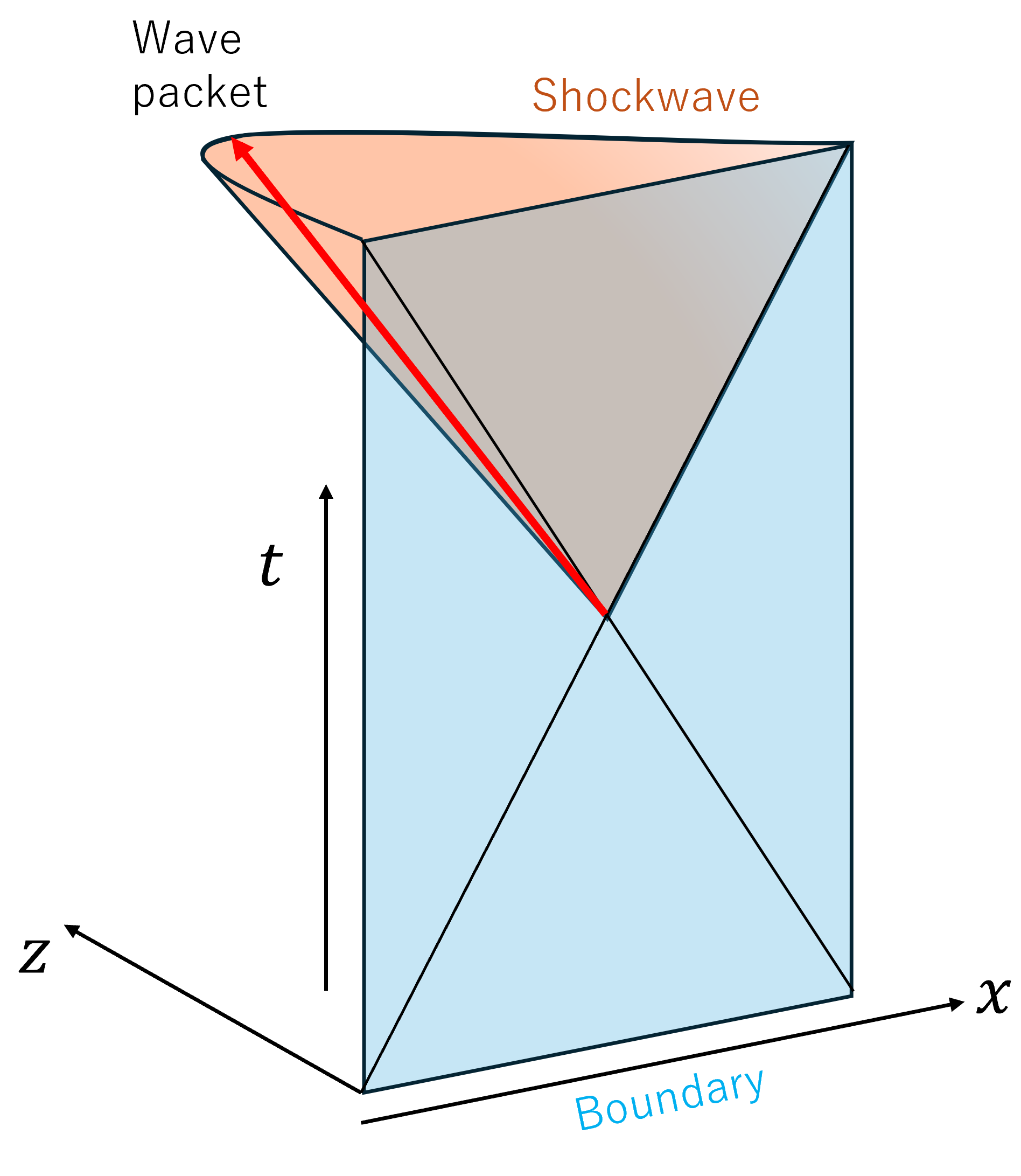}
    \caption{A wave packet propagating from the origin on the boundary into the bulk, and the associated spherically symmetric shockwave. }
    \label{f10}
\end{figure}

This expectation value is similar to the zero-temperature result \eqref{expO0}.
The essential difference is that $1/u^2$ is replaced by $1/(\beta^2 \sinh^2 (\pi u/\beta))$.
Because of this difference, the expectation value exponentially decays for large $u$ or $v$ along the light cone.
Away from the light cone, the expectation decays like 
\begin{align}
    e^{-\frac{2 \pi}{\beta} (|u|+|v|)} = 
    \begin{cases}
    e^{-\frac{4 \pi}{\beta} |t|} & \text{for } |t| > |x|, \\
    e^{-\frac{4 \pi}{\beta} |x|} & \text{for } |x| > |t|,
    \end{cases}
\end{align}
for large $|t|$.
For $t=t_0$ with a large fixed $t_0$, this is constant inside the light cone, i.e., $|t_0| > |x|$, and exponentially decays, $e^{-\frac{4 \pi}{\beta} (t_0 +(|x|-t_0))}$, outside the light cone. 
This uniform distribution in this region could be a form of thermalization.\footnote{
For a two-dimensional CFT, the energy density can not diffuse because of the conservation law.
For the holographic CFT, this is because the (boundary) gravitons should be on the boundary in the bulk picture. 
Thus, this is not a usual sense of the thermalization process.
}
Note that our computations are exact and universal for any two-dimensional CFT, not restricted to holographic CFT.
Thus, this characteristic plateau-like behavior of the expectation value for the localized excited state by a primary operator should appear in any CFT.

One important thing here is that, while the localization on the light cone is similar to the zero-temperature case, the late-time decay is exponential rather than power-law, and it depends explicitly on the temperature. 
Thus, the expectation value shows ``diffusion'' or ``dissipation'' as a finite-temperature effect. 

\begin{figure}
    \centering
    \includegraphics[width=0.8\linewidth]{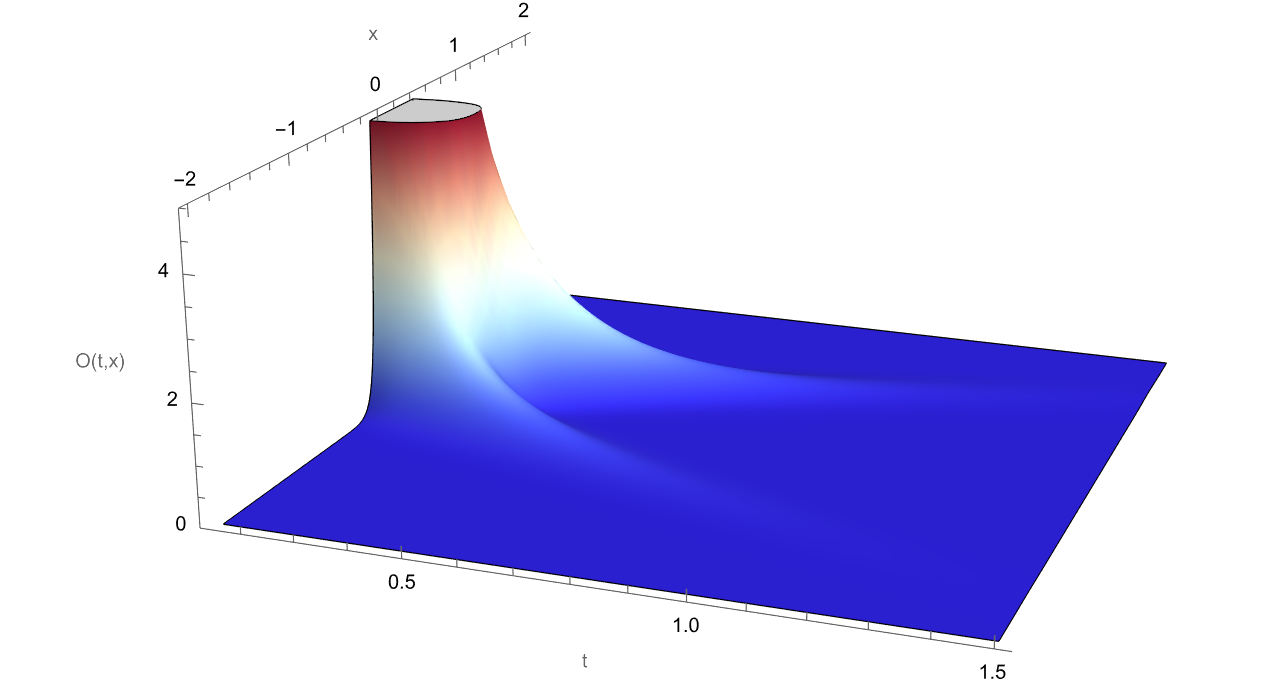}
    \caption{
    \md{ Profile of \(\langle\cO(t,x) \rangle_{\beta,\mathrm{wp}}\) given by \eqref{evo} for \(a=0.15, p_u =10, p_v = 25, \beta = 3, \ep = 0.1\).}
    \md{The contribution from the fourth term, which spreads spatially, is neglected.}
    Shockwave-induced contributions decay exponentially along the light cone at finite temperature. In the high-temperature regime with \(\beta p_u \sim \cO(1)\), the decay is excessively rapid. For this reason, we set \(\beta=3\) in the present plot.}
    \label{f3}
\end{figure}

Next, we will consider $\zeta \sim {\cal O}(1)$ case.
For this case, we can show that the summation over $n$ and $m$
in \eqref{F1} is reduced to $n=0$ and $m=0$ if we neglect $e^{-\alpha (p_u a)^2 }$ where $\alpha ={\cal O}(1)$ positive constant.
Indeed, $\zeta$ and $n$ in $e^{-\frac{(u+i\beta(n-\zeta))^2}{4a^2}}$ are of same order and then, 
the contribution for 
$n>0$ is proportional to $e^{\frac{(\beta(n-\zeta))^2}{4a^2}}=e^{\frac{-\beta^2 n (2 \zeta-n))}{4a^2}} e^{\frac{(\beta \zeta)^2}{4a^2}}$, which is 
smaller than
$e^{\frac{(\beta \zeta)^2}{4a^2}}=
e^{(p_u a)^2/4}$
for $n=0$ by a factor of $e^{-\alpha (p_u a)^2 }$.
For the $u_3$ integration, there are contributions from the poles of the $\sinh$ function, but only the pole at $u_3=u$ should be kept in our approximation.
Therefore, the results for this case are the same as the results for the $\zeta \ll 1$ case.
Figure~\ref{f3} shows the distribution of the expectation value \(\langle \cO(t,x)\rangle_{\beta,\mathrm{wp}}\).


Finally, we will consider the $\zeta \gg 1$ case.
In this case, we should retain contributions from the various poles.
We leave the details of the calculation to Appendix~\ref{zetagg}  and show only the results here.
We obtained the following expression:
\begin{align}
    \langle\cO(t,x) \rangle_{\beta,\mathrm{wp}}
    &\simeq 2C_{123} \Bigg(
    e^{-\frac{u^2}{2a^2}}e^{-\frac{v^2}{2a^2}}\frac{\pi}{a^2}\frac{1}{1-e^{-\frac{\beta p_u}{2}}}\frac{1}{1-e^{-\frac{\beta p_v}{2}}}
    + e^{-\frac{u^2}{2a^2}}\frac{\pi^2}{ap_v\beta^2}\sqrt{2\pi}\frac{1}{\sinh^2\frac{\pi v}{\beta}}\frac{1}{1-e^{-\frac{\beta p_u}{2}}}\nn
    & \hphantom{\simeq2C_{123}}
    + e^{-\frac{v^2}{2a^2}}\frac{\pi^2}{ap_u\beta^2}\sqrt{2\pi}\frac{1}{\sinh^2\frac{\pi u}{\beta}}\frac{1}{1-e^{-\frac{\beta p_v}{2}}}
    +\frac{2\pi^4}{\beta^4 p_up_v}\frac{1}{\sinh^2\frac{\pi u}{\beta}\sinh^2\frac{\pi v}{\beta}}
    \Bigg).\label{expotg}
\end{align}
For these analyses, the assumptions concerning the relevant parameters are \(\zeta \coloneqq \frac{a^2p_u}{\beta} \gg 1\), \(\xi \coloneqq \frac{a^2p_v}{\beta} \gg 1\), and \(ap_u,ap_v \gg 1\).
\MD{The first two assumptions correspond to the high-temperature regime, while the last two correspond to the wave packet being well-localized in the bulk.}
The result \eqref{expotg} differs from the result \eqref{evo} for the $\zeta \ll 1$ case only by the $e^{-\beta p/2}$ factors.
The case where these factors are not small corresponds to the high temperature case with $a/\beta \gg 1$, since we assumed  $ a p_u \gg 1$.
In this case,
the wave packet is much larger than the typical black hole scale, and the wave packet is affected by the curvature effects even at $t=0$.
Because of this, interpretations of the deformation by $e^{-\beta p/2}$ factors will not be simple.

\subsection{Energy density}\label{energy}

In~\cite{Terashima:2023mcr,Tanahashi:2025fqi}, the energy density in the CFT was computed for wave packets propagating in pure AdS. In this section, we evaluate the energy density in a two-dimensional finite-temperature CFT, corresponding to the case with a bulk black hole. Due to the special features of two-dimensional CFTs, we find that thermal effects may not appear in the energy density without
explicit computations.

\subsection*{Zero-temperature}
\begin{figure}[htbp]
  \begin{minipage}[t]{0.45\linewidth}
    \centering
    \includegraphics[width=5cm]{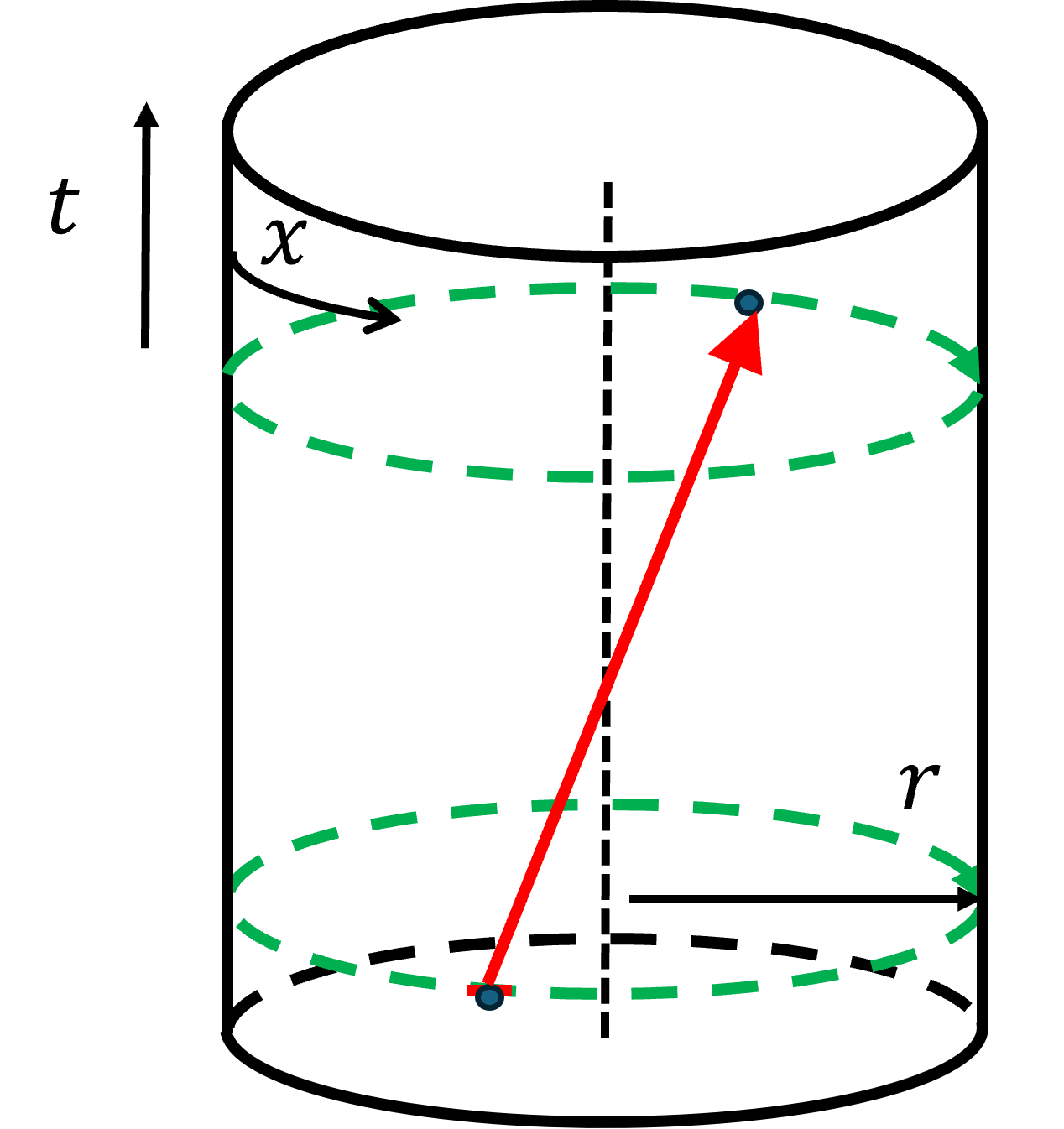}
    \caption{A wave packet propagating from the boundary to the boundary.}
    \label{f4}
  \end{minipage}
\hspace{10mm} 
  \begin{minipage}[t]{0.45\linewidth}
    \centering
    \includegraphics[width=4.8cm]{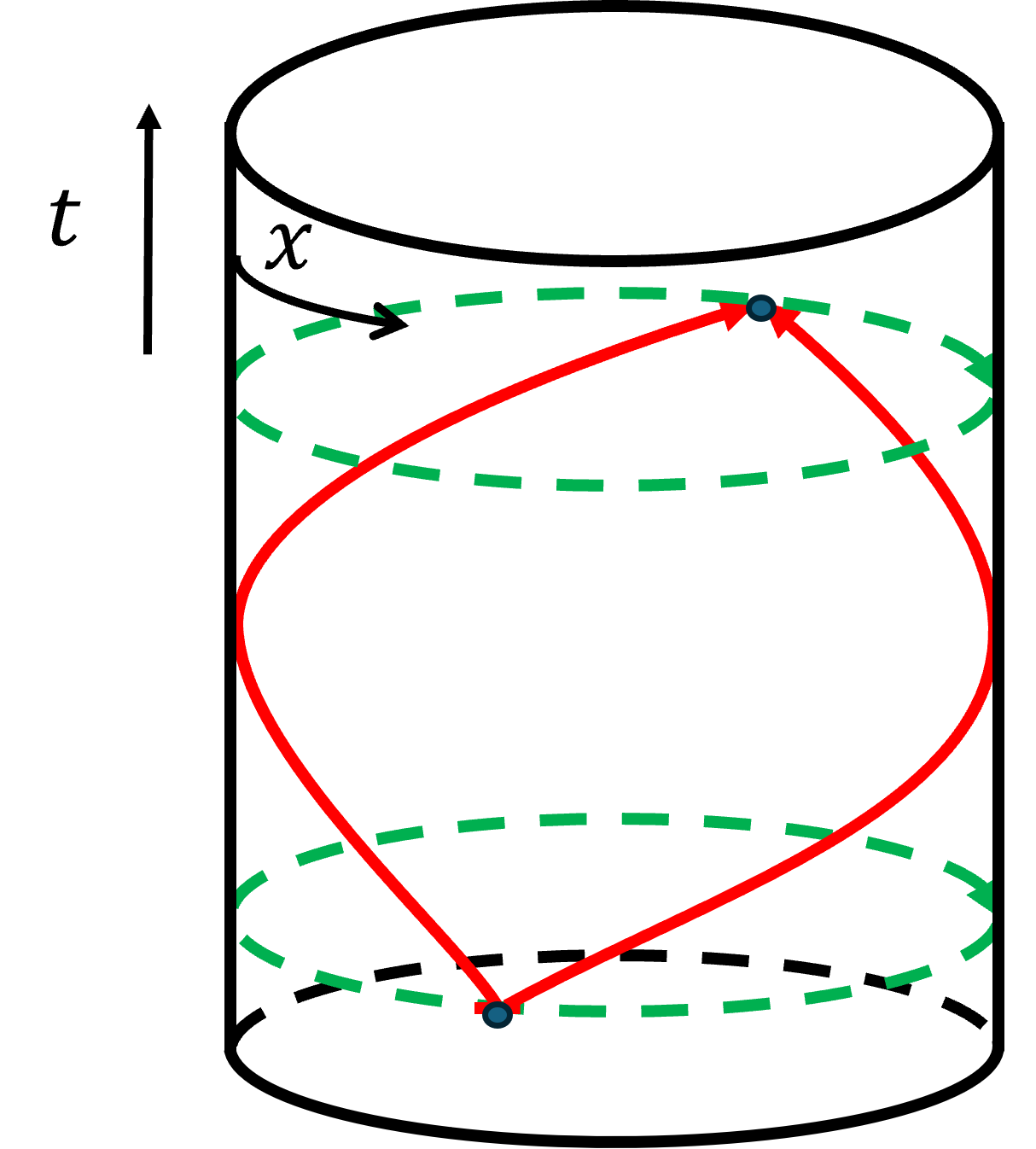}
    \caption{Two ``particles'' on the CFT side corresponding to the bulk wave packet.}
    \label{f5}
  \end{minipage}
\end{figure}

We begin by reviewing the energy density in the two-dimensional zero-temperature CFT corresponding to a wave packet propagating in pure AdS\(_3\), as studied in previous work~\cite{Terashima:2023mcr}. In this paper, we compute the sub-leading contributions that were neglected in that analysis.
The energy density  is computed as
\begin{align}
   & \bra{p,\omega}   T_{00}(t, x) \ket{p,\omega} \nn
   =
   & \int d t_1 \, d x_1 \, e^{-\frac{ (x_1)^2+t_1^2}{2 a^2}-i p x_1+i {\omega} t_1} 
   \int d t_3 \, d x_3 \, e^{-\frac{ (x_3)^2+t_3^2}{2 a^2}+i p x_3-i {\omega} t_3}  \nn
   & \times \frac{\Delta}{2} 
     \left(
      \frac{(u_1-u_3)^2}{(u-u_1)^2(u-u_3)^2}
      +
        \frac{(v_1-v_3)^2}{(v-v_1)^2(v-v_3)^2}
      \right)
    \frac{1}{(u_1-u_3)^{ \Delta} (v_3-v_1)^{ \Delta} } \nn
    =
    & \frac{1}{4}\int d u_1 \, d v_1 \, d u_3 \, d v_3 \,\, e^{-\frac{ (u_1-i p_u a^2)^2+(v_1-i p_v a^2)^2+(u_3+i p_u a^2)^2+(v_3+i p_v a^2)^2}{4 a^2} 
    - \frac{a^2}{2} ((p_u)^2+(p_v)^2) }
     \nn
   & \times \frac{\Delta}{2} 
     \left(
      \frac{(u_1-u_3)^2}{(u-u_1)^2(u-u_3)^2}
      +
        \frac{(v_1-v_3)^2}{(v-v_1)^2(v-v_3)^2}
      \right)
    \frac{1}{(u_1-u_3)^{ \Delta} (v_3-v_1)^{ \Delta} },
    \label{T00_AdS3CFT2}
\end{align}
where we defined \(p_u \coloneqq \omega - p \,\, (\gg 0)\) and \(p_v \coloneqq \omega + p \,\, (\gg 0)\). We refer the reader to the previous work~\cite{Terashima:2023mcr} for further details.
The resulting energy density, including the normalization constant, is given by the following expression:
\begin{align}
   {\cal E}(t,x) = &  \frac{1}{{\cal N}_0^2} \, \bra{p,\omega} \frac{1}{2 \pi}  T_{00}(t, x) \ket{p,\omega} \nn
      \simeq & 
  \frac{1}{2 \sqrt{2 \pi} a}
  \left( e^{-\frac{ u^2}{2 a^2}  } p_u + e^{-\frac{ v^2}{2 a^2}  } p_v \right).
  \label{calE_AdS3CFT2}
\end{align}
The energy in the vicinity of \(u=0\) and \(v=0\) is found to be \(p_u/2\) and \(p_v/2\), respectively. Adding these contributions yields the total energy \(\omega\), consistent with the wave-packet state under consideration. This result shows that, in the CFT, energy propagates at the speed of light, which is causally consistent with the picture of a wave packet propagating through the bulk as shown in Figures~\ref{f4} and \ref{f5}.\footnote{
This calculation was performed in the Poincaré patch. These figures are a naive extension into the global patch.
We also note that these pictures are more accurate for the state given by acting with unitary wave-packet operators on the vacuum, as discussed in~\cite{Terashima:2023mcr} and Appendix~\ref{unt}.
}
\MD{
When extended to the global patch, the time it takes for a wave packet sent from the boundary to propagate through the bulk and arrive at the antipodal point on the boundary is expected to coincide with the time in the CFT at which a split energy pulse merges again.
}
\\
In~\cite{Terashima:2023mcr}, the contribution from the pole at \(u_1=u_3\) is neglected, since the \(u_3\)-integration becomes independent of \(p_u\).
As an example, for \(\Delta=3\),\footnote{
\(\Delta = 3\) is the simplest integer value that preserves the pole structure. In contrast, for \(\Delta=2\), the pole at \(u_1 =u_3\) is absent.
} the following contribution arises from a calculation on \(u_1=u_3\) pole similar to that for the expectation value of \(\cO\):
\begin{align}
      -\frac{6}{\pi p_u^2 u^4}-\frac{6}{\pi p_v^2 v^4}.
      \label{energy-density_zeroT-part}
\end{align}
\MD{Note that since the saddle-point approximation was performed under the assumption \(|u|,|v|\gg a\), the result does not become singular. }
 Although these contributions are not important in the present context, we compute them here for the sake of comparison with the finite-temperature case in Appendix~\ref{ded}. 

\subsection*{Finite-temperature}
We now consider the finite-temperature case.\footnote{
For the localized excited state, the expectation value of the energy-momentum tensor was computed in~\cite{Caputa:2014eta}.
} 
In two-dimensional conformal field theory, when written in complex coordinates, the energy-momentum tensor splits into holomorphic and anti-holomorphic components. Specifically, the components \(T(z)\) and \(\bar{T}(\bar{z})\) are independent and capture the left- and right-moving sectors of the theory. By a conformal transformation, the zero-temperature CFT on the plane is mapped to a finite-temperature CFT on the cylinder. The energy-momentum tensor transforms as follows:
\begin{align}
    T_{\textrm{cyl.}}(w) = \left(\frac{2\pi}{\beta}\right)^2 \left\{ T_\textrm{pl.}(z) z^2 - \frac{c}{24} \right\},
\end{align}
where \(c\) denotes the central charge. Combining the transformation of the primary operators and the energy-momentum tensor, we obtain the following expression for the three-point function on the cylinder:
\begin{align}
    &\langle \cO(w_1,\bar{w}_1) T(w)\cO(w_3,\bar{w}_3)\rangle_\beta  \nn
    = &  \left( \frac{2\pi}{\beta }\right)^2  \left( \frac{2\pi}{\beta}\right)^{2\Delta} 
    \abs{z_1}^\Delta \abs{z_3}^\Delta \left\{ \langle \NT{z^2}\cO(z_1,\bar{z}_1) T(z)\cO(z_3,\bar{z}_3)\rangle
    -\frac{c}{24} \langle \cO(z_1,\bar{z}_1)\cO(z_3,\bar{z}_3)\rangle  \right\} \nn
    = &  \left( \frac{2\pi}{\beta }\right)^2  \left( \frac{2\pi}{\beta}\right)^{2\Delta} 
    \abs{z_1}^\Delta \abs{z_3}^\Delta \nn
    & \times\left\{ \sum_{i=1,3}\left( \frac{\Delta/2}{(z-z_i)^2}+\frac{\partial_i}{z-z_i}\right)z^2\langle \cO(z_1,\bar{z}_1)\cO(z_3,\bar{z}_3)\rangle 
    -\frac{c}{24} \langle \cO(z_1,\bar{z}_1)\cO(z_3,\bar{z}_3)\rangle  \right\} \nn
    = &  \left( \frac{2\pi}{\beta }\right)^2  \left( \frac{2\pi}{\beta}\right)^{2\Delta} 
    \abs{z_1}^\Delta \abs{z_3}^\Delta \nn
    & \times\left\{ \frac{\Delta}{2}\left( \frac{(z_1-z_3)^2}{(z-z_1)^2(z-z_3)^2}\right)z^2\langle \cO(z_1,\bar{z}_1)\cO(z_3,\bar{z}_3)\rangle 
    -\frac{c}{24} \langle \cO(z_1,\bar{z}_1)\cO(z_3,\bar{z}_3)\rangle  \right\} \nn
    = &  \left( \frac{\pi}{\beta }\right)^2  \left( \frac{\pi}{\beta}\right)^{2\Delta} 
       \frac{\Delta}{2}
    \frac{\sinh^2 \frac{\pi(w_1-w_3)}{\beta}}{\sinh^2 \frac{\pi(w-w_1)}{\beta}\sinh^2 \frac{\pi(w-w_3)}{\beta}}
    \frac{1}{ \sinh^\Delta \frac{\pi(w_1-w_3)}{\beta}\sinh^\Delta \frac{\pi(\bar{w}_1-\bar{w}_3)}{\beta}}\nn
    & -\frac{c}{24}\left( \frac{2\pi}{\beta }\right)^2  \left( \frac{2\pi}{\beta}\right)^{2\Delta} 
        e^{\frac{\pi\Delta (w_1 + \bar{w}_1)}{\beta}}e^{\frac{\pi\Delta (w_3 + \bar{w}_3)}{\beta}}\langle0| \cO(z_1,\bar{z}_1)\cO(z_3,\bar{z}_3)|0\rangle.
    \label{OTO_beta}
\end{align}
The second term is canceled by the normalization and gives rise to a constant term 
\(-\frac{c \pi^2}{6\beta^2}\). The anti-holomorphic contribution can be obtained by replacing \(w_i\leftrightarrow \bar{w}_i\), and the full energy density includes both contributions, resulting in an overall factor of 2. In the following, we denote the first term as \(A\) and focus on its contribution,  which corresponds to the part localized at \(u=0\). Almost all of the calculations for the energy density follow the same steps as those for the expectation value of \(\cO\). The only difference lies in the separation between the \(u\)- and \(v\)-components, which leads to a significant distinction in the result.

As in the previous sections, we perform analytic continuation together with the \(i\ep\)-prescription, and integrate the result with wave packet smearing. Specifically, we make the replacements \(w\rightarrow u, ~ \bar{w} \rightarrow -v,\) and \(t_1 \rightarrow t_1 -i\ep, ~ t_3 \rightarrow t_3 + i\ep\).\\
What we need to compute is the following:
\begin{align}
&\langle A\rangle_{\beta,\mathrm{wp}} \nn
    \coloneqq&  \frac{1}{4}\int du_1 dv_1 du_3 dv_3 e^{-\frac{(u_1-ip_u a^2)^2 + (v_1-ip_va^2)^2+(u_3+ip_ua^2)^2+(v_3+ip_va^2)^2}{4a^2}
    -\frac{a^2p_u^2}{2}-\frac{a^2p_v^2}{2}}A_\beta\nn
    =&  \frac{1}{4}\int du_1 dv_1 du_3 dv_3 e^{-\frac{(u_1-ip_u a^2)^2 + (v_1-ip_va^2)^2+(u_3+ip_ua^2)^2+(v_3+ip_va^2)^2}{4a^2}
    -\frac{a^2p_u^2}{2}-\frac{a^2p_v^2}{2}} \nn
    & \times  \left( \frac{\pi}{\beta }\right)^2  \left( \frac{\pi}{\beta}\right)^{2\Delta} 
       \frac{\Delta}{2}
    \frac{\sinh^2 \frac{\pi(u_1-u_3-i\ep)}{\beta}}{\sinh^2 \frac{\pi(u-u_1+i\ep)}{\beta}\sinh^2 \frac{\pi(u-u_3-i\ep)}{\beta}}
    \frac{1}{ \sinh^\Delta \frac{\pi(u_1-u_3-i\ep)}{\beta}\sinh^\Delta \frac{\pi(v_3-v_1+i\ep)}{\beta}}.
    \label{A_beta-def}
\end{align}
The detailed calculation is not essential here and is deferred to Appendix~\ref{ded}.
What is important is that
\NTmod{the expectation value of \(T(u)\) does not depend on $v$ by definition \eqref{OTO_beta} and \eqref{A_beta-def}, and also the $p_v$ dependence is canceled out by dividing it by the normalization factor~\(\cN_\beta^2\).}
Consequently, unlike the case of the expectation value of \(\cO\), the energy density can be approximately expressed 
\NTmod{in a form factorized into the $u$ and $v$-dependent part}
as
\begin{align}
    \mathcal{E}(t,x) 
    \NTmod{
    =\frac{
\bra{p,\omega} \frac{-1}{2 \pi} \bigl(T(u) + \bar T(v)\bigr) \ket{p,\omega}    
    }{{\cal N}_\beta^2}
    }
    \simeq  \mathrm{const.} \times \lr{p_ue^{-\frac{u^2}{2a^2}}F(\beta p_u)+p_ve^{-\frac{v^2}{2a^2}}G(\beta p_v)}+\frac{c\pi^2}{3\beta^2},
    \label{energy-density_finite-T}
\end{align}
where \(F\) is independent of \(v\) and \(G\) is independent of \(u\).\footnote{
To be precise, both of them depend also on \(a\).
}

Let us recall the expectation value of the scalar discussed in section~\ref{ftcft}.
Away from the origin, \(u\) and \(v\) cannot be simultaneously zero, and as a result, the expectation value of \(\cO\) exhibits thermal suppression through cross terms involving both \(u\)- and \(v\)-dependent contributions.
In contrast, in two-dimensional CFT, the energy density naturally separates into independent \(u\)- and \(v\)-dependent components, and thus no thermal suppression can appear along the light cone, along which the energy propagates. Furthermore, when a finite temperature is introduced, new poles appear along the imaginary axis. However, since their real parts remain unchanged, the localization of energy on the light cone is not affected.
This implies that, even if the wave packet experiences a delay due to the presence of a black hole in the bulk, the propagation of the energy density on the CFT side does not exhibit any corresponding delay.

This can be understood from the fact that $T_{00}\sim T(z)+ T(\bar{z})$, which implies that 
$\mathcal{E}(t,x) = f(t-x)+g(t+x)$.
Thus, the energy density at $t=t_0$ is given by 
the shift of $x$ by $\pm t_0$ of 
the initial energy density 
$\mathcal{E}(t=0,x) = f(-x)+g(x)$.
Note that this property holds for the CFT on $S^1$, which corresponds to the BTZ black hole case.
If the wave packet can reach another boundary point of the BTZ black hole, the delay in the black hole geometry, compared with the AdS geometry, as the Gao–Wald theorem~\cite{Gao:2000ga} indicated, will contradict this light-like behavior in the CFT picture.
However, there are no bulk light-like geodesics in the BTZ black hole geometry that connect two asymptotic boundary points, so there is no contradiction.
It is interesting to see such a delay by the finite temperature effects in the CFT picture for the higher-dimensional black hole case, as claimed in~\cite{Terashima:2021klf}.

\section{Bulk picture 
}
\label{sec:discussion}

In principle, we can compute the expectation values of the states in the bulk picture using the in-in formalism in perturbation theory, which involves the retarded Green function~\cite{Satoh:2002bc, Balasubramanian:2012tu} and the density given by the absolute value of the ``wave function''.
Here,
instead of making a detailed analysis using these, we will discuss some general aspects.

First, we will discuss the scalar expectation value for the wave packet state given in \eqref{expO0} for zero temperature. 
In particular, for $|u| \gg a, \, |v| \gg a $, it behaves as
\begin{align}
    \langle \cO(t,x)\rangle_{0,\mathrm{wp}}
    \simeq  \frac{4 C_{123}}{p_up_v}
    \frac{1}{\lr{u}^2\lr{v}^2},
    \label{expOB}
\end{align}
for $\Delta=2$, and we find that 
\begin{align}
    \langle \cO(t,x) \rangle_{0,\mathrm{wp}}
    \sim \frac{ C_{123}}
    {(p_up_v)^{\frac{\Delta}{2}}} \, 
    \frac{1}{\lr{u v}^\Delta},
    \label{expOB2}
\end{align}
for general $\Delta$ as shown in \eqref{generalD}, which is the same form as the one for the localized excited state discussed in Appendix~\ref{shws}.
We note that this is proportional to the two-point function, although we do not think there is any direct relation between the two-point function and the expectation value.
Indeed, for another primary scalar $\tilde{\cO}$ that has a vanishing two-point function with $\cO$, we can compute $\langle \tilde{\cO}(t,x)\rangle_{0,\mathrm{wp}}$, then it becomes \eqref{expOB2} if $C_{123} \neq 0$.

By this, the expectation value of the bulk scalar in the leading order perturbation is given by
\begin{align}
    \langle \phi(t,x,z)\rangle_{0,\mathrm{wp}}
    \sim C_{123}
    \frac{1}{\lr{z^2-u v}^\Delta},
    \label{expp}
\end{align}
because this is the solution of the bulk free equation of motion and the boundary value for $z\rightarrow 0$ is given by \eqref{expOB2}. This is, of course, the bulk-boundary propagator, although the interpretation will not be straightforward again.
For $\Delta \gg 1$, this is localized near the bulk light cone of $z=u=v=0$, which is the shockwave-like distribution shown in Fig.~\ref{f10}.
This shockwave is associated with the bulk-propagating wave packet, and is consistent with the previous result in~\cite{Afkhami-Jeddi:2017rmx}.\footnote{In Appendix~\ref{shws}, we present the calculation for the case of the shockwave state.
} 
For $\Delta={\cal O}(1)$, it is not localized, because of the entanglement in the vacuum. Indeed, if we use the unitary wave packet operators discussed in Appendix~\ref{unt}, this term is absent.

The other terms in \eqref{expO0} are localized at $u={\cal O}(a)$ and/or $v={\cal O}(a)$, and then, we expect that the bulk distributions of the expectation values corresponding to these terms come from the wave packet on the null geodesic and are given by the shockwave-like distribution shown in Fig.~\ref{f10}.

Next, we will discuss the scalar expectation value for the wave packet state given in \eqref{evo} for the finite temperature. 
For $|u| \gg a, \, |v| \gg a $, we expect that
\begin{align}
    \langle\cO(t,x) \rangle_{\beta,\mathrm{wp}} \sim \frac{1}{\sinh^\Delta\frac{\pi u}{\beta}\sinh^\Delta\frac{\pi v}{\beta}}.
    \label{expotg1}
\end{align}
for general $\Delta$, which is the same form as the one for the localized excited state again.
We note that this is proportional to the two-point function for the finite temperature.
Thus, the bulk expectation value, $ \langle \phi(t,x,z)\rangle_{\beta,\mathrm{wp}} $, is given by 
the bulk-boundary propagator for the finite temperature.

\begin{figure}
    \centering
    \includegraphics[width=0.5\linewidth]{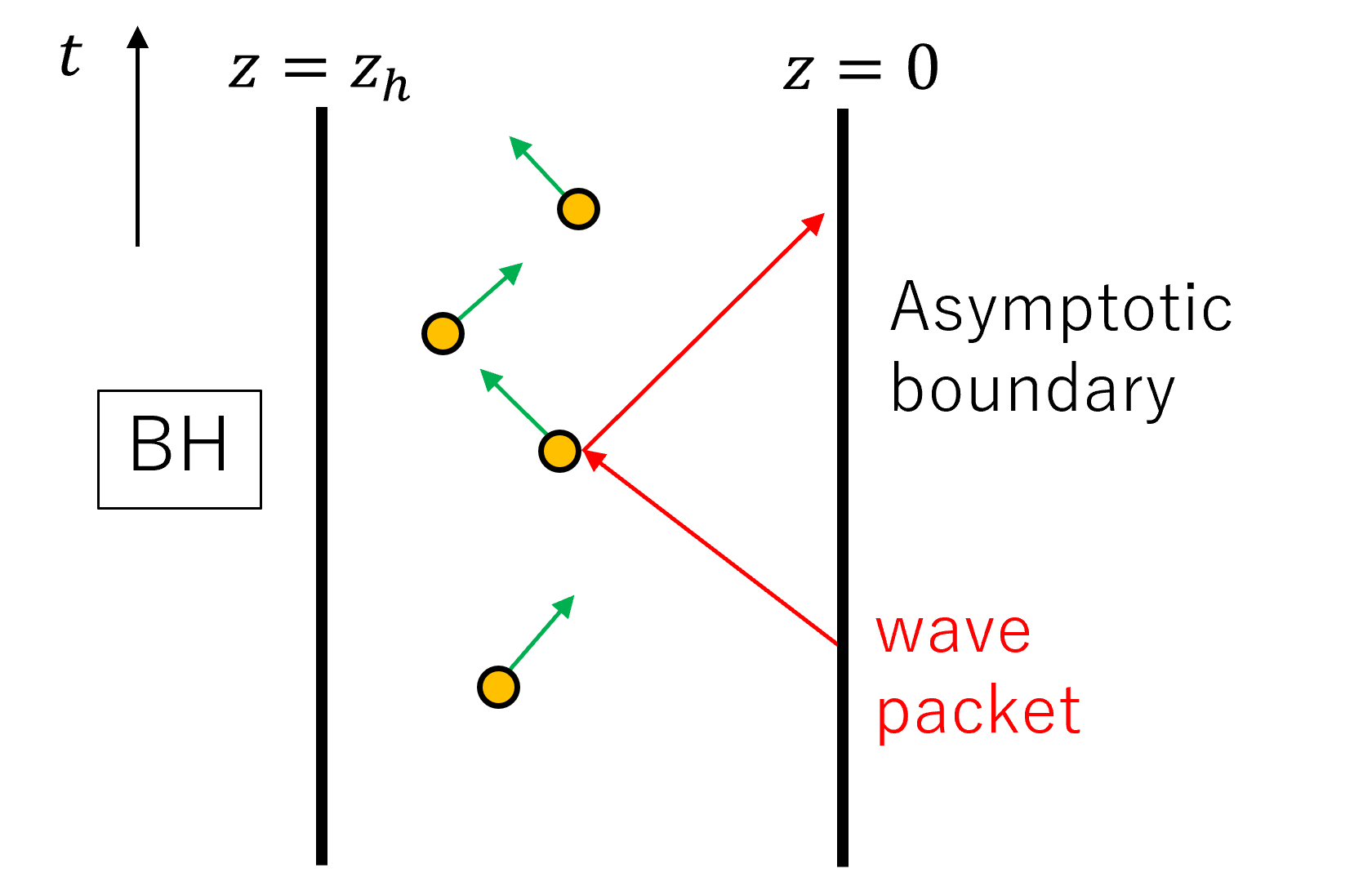}
    \caption{In the presence of a black hole, the wave packet can be scattered by thermal excitations at inverse temperature \(\beta\), and may return to the boundary after the scattering. The orange circles represent thermal excitations with energy of order the temperature \(\beta^{-1}\). The vertical line on the left indicates the black hole horizon, while the vertical line on the right denotes the asymptotic boundary.}
    \label{f6}
\end{figure}

It should be noted that the expectation value \eqref{expotg1} (and its bulk counterpart) is not localized on the light cone even for $\Delta \gg 1$.
In particular, for $|u|, |v| \gg \beta$, it approximately is given by 
\begin{align}
    e^{-\frac{\Delta \pi}{\beta} (|u|+|v|)} = 
    \begin{cases}
    e^{-\frac{2 \Delta \pi}{\beta} |t|} & \text{for } |t| > |x|, \\
    e^{-\frac{2 \Delta \pi}{\beta} |x|} & \text{for } |x| > |t|.
    \end{cases}
\end{align}
This may be explained by the finite temperature effect.
In the bulk picture, the Green function will be an analytic continuation of the Euclidean Green function for the Hartle-Hawking vacuum. 
This includes the effect of the thermal radiation around the black hole as shown in Fig.~\ref{f6}.

\paragraph*{\re{Bulk perturbative computations}}

Here, we will briefly explain the bulk perturbative computations of the expectation values at zero temperature \blue{and finite temperature. First, we will consider the zero temperature case.} 
We denote bulk Heisenberg operator as $\phi(t, x, z)$ and 
consider the wave packet state \eqref{wads},
\begin{align}
     \ket{p, \omega} 
    &= \lim_{z \to 0} \frac{1}{z^\Delta} 
    \int dt \, d x \, e^{-\frac{t^2+x^2}{2a^2}- i \omega t+ip x} \phi(t, x, z) \ket{0}.
\end{align}
Then, the expectation value in the Heisenberg picture is given by
\begin{align}
     \langle \phi(t,x,z) \rangle_{0,\mathrm{wp}}= \bra{p, \omega} \phi(t, x, z) \ket{p, \omega} =\bra{p, \omega} e^{iHt} \phi(t=0, x, z) e^{-iHt}\ket{p, \omega},
\end{align}
where $H$ is the full Hamiltonian.\footnote{We assumed that the CFT primary operator is given by the full bulk local field via the BDHM relation~\cite{Banks:1998dd}.}
Now, we decompose the Hamiltonian as $H=H_0+ \lambda H_I$, where $H_0$ is the free Hamiltonian, and
we find 
\begin{align}
      \phi(t, x, z)= e^{iHt} \phi(t=0, x, z) e^{-iH t}= e^{iH_0 t} U \phi(t=0, x, z) U^\dagger e^{-iH_0 t},
\end{align}
where
\begin{align}
      U = \md{\mathcal T}\exp \left( \int_0^1 ds  e^{-i H_0 t s} (-i \lambda H_I t) e^{i H_0 t s} \right) = 1-i \lambda  \int_0^t d t' \, e^{-i H_0 t'} H_I e^{i H_0 t'} +\cdots.
\end{align}
Then, in the leading order in $\lambda$, we obtain
\begin{align}
     \phi(t,x,z) &\simeq  e^{iH_0 t} \phi(t=0,x, z) e^{-iH_0 t}
     -i \lambda  e^{iH_0 t} 
     \int_0^t d t' \, [e^{-i H_0 t'} H_I e^{i H_0 t'}, \,
      \phi(t=0, x, z)] e^{-iH_0 t} \nn 
      &=\phi_0(t,x, z) 
     -i \lambda 
     \int_0^t d t' \, [H_I(t'), \,
      \phi_0(t, x, z)] ,
\end{align}
where we define $\phi_0(t,x, z) =e^{iH_0 t} \phi(t=0, x, z) e^{-iH_0 t}$ and $H_I(t')=e^{i H_0 (t')} H_I e^{-i H_0 (t')}$.
This implies that our wave packet state in the bulk perturbation theory should be represented as 
\begin{align}
     \ket{p, \omega} 
    &\simeq  \lim_{z \to 0} \frac{1}{z^\Delta} 
    \int dt \, d x \, e^{-\frac{t^2+x^2}{2a^2}- i \omega t+ip x} \left( \phi_0(t,x, z) 
     -i \lambda 
     \int_0^t d t' \, [H_I(t'), \,
      \phi_0(t, x, z)] \right) \ket{0},
\end{align}
where the interaction gives the multi-particle state,
and the expectation value is given by
\begin{align}
     \langle \phi(t,x,z) \rangle_{0,\mathrm{wp}} & \simeq \bra{p, \omega}  \phi_0 (t, x, z) \ket{p, \omega} 
     -i \lambda \bra{p, \omega} \int_0^t d t' \, [H_I(t'), \,
      \phi_0(t, x, z)] \ket{p, \omega}, \nn
      & \simeq \lim_{z_1 \to 0} \frac{1}{z_1^\Delta} 
    \int dt_1 \, d x_1 \, e^{-\frac{t_1^2+x_1^2}{2a^2} + i \omega t_1 - ip x_1} \lim_{z_3 \to 0} \frac{1}{z_3^\Delta} 
    \int dt_3 \, d x_3 \, e^{-\frac{t_3^2+x_3^2}{2a^2} - i \omega t_3 + ip x_3} \nn
    & \times (-i \lambda) \bra{0} \left( \phi_0(t_1, z_1, x_1)  \int_0^t d t' \, [H_I(t'), \,
      \phi_0(t, x, z)] \phi_0(t_3, z_3, x_3) \right. \nn
      & \hspace{2cm} \left. 
       + \int_0^{t_1} d t' \, [H_I(t'), \, 
      \phi_0(t_1, z_1, x_1)] \phi_0(t, x, z) \phi_0(t_3, z_3, x_3) \right. \nn
      & \hspace{2cm} \left. 
       +\phi_0(t_1, z_1, x_1) \phi_0(t, x, z) \int_0^{t_3} d t' \, [H_I(t'), \,
      \phi_0(t_3, z_3, x_3)] 
      \right) \ket{0},
\end{align}
where ${\cal O}(\lambda^0)$ terms vanished because the free theory does not have a nonzero three-point function.

Here, the bulk metric is 
\[
ds^2=\frac{L^2}{z^2}\big(-dt^2+dx^2+dz^2\big),\qquad z>0,\quad x\in\mathbb R,
\]
and the action for the bulk scalar is
\[
S=\int dtdx dz \sqrt{g} \left( \tfrac12 g^{\mu\nu}\nabla_\mu\phi\,\nabla_\nu\phi-\tfrac12 m^2\phi^2-\frac{\lambda}{3!}\phi^3 \right), 
\]
where \(\nu = \sqrt{1+m^2L^2}\) and \(\Delta\equiv 1+\nu\).
The interaction Hamiltonian is 
\[
H_I(t) =\int dx dz \sqrt{g} \frac{1}{3!}(\phi_0(t,x,z))^3,
\]
and 
\begin{align}
    -\int_0^t d t' \, [H_I(t'), \,
      \phi_0(t, x, z)] =\int_0^t d t' \int dx' dz' \sqrt{g} \frac{1}{2}(\phi_0(t',x',z'))^2  [\phi_0(t, x, z), \phi_0(t',x', z')],
\end{align}
where $[\phi_0(t, x, z), \phi_0(t',x', z')]$ is a constant and proportional to the retarded propagator \(\Delta_R\) because $t \geq t'$.
More precisely, the $(\phi_0)^3$ in the interaction Hamiltonian should be taken as the normal-ordered products, and $(\phi_0)^2$ in the above equation is also a normal-ordered product.
Then, the expectation value is given by
\begin{align}
     \langle \phi(t,x,z) \rangle_{0,\mathrm{wp}} 
      & \simeq  (i \lambda) \lim_{z_1 \to 0} \frac{1}{z_1^\Delta} 
    \int dt_1 \, d x_1 \, e^{-\frac{t_1^2+x_1^2}{2a^2} + i \omega t_1 - ip x_1} \lim_{z_3 \to 0} \frac{1}{z_3^\Delta} 
    \int dt_3 \, d x_3 \, e^{-\frac{t_3^2+x_3^2}{2a^2} - i \omega t_3 + ip x_3} \nn
    & \times  \left( \int_0^t d t' \int dx' dz' \sqrt{g} [\phi_0(t, x, z), \phi_0(t',x', z')] D^+(X_1,X') D^+(X',X_3) \right. \nn
    &+\int_0^{t_1} d t' \int dx' dz' \sqrt{g} [\phi_0(t_1, x_1, z_1), \phi_0(t',x', z')] D^+(X',X) D^+(X',X_3) \nn
    &+ \left. \int_0^{t_3} d t' \int dx' dz' \sqrt{g} [\phi_0(t_3, x_3, z_3), \phi_0(t',x', z')] D^+(X_1,X') D^+(X,X') \right),
    \label{vev}
\end{align}
where 
\begin{align}
    D^+(X,X')=\bra{0} \phi_0(t, z, x) \phi_0(t',x',z')\ket{0},
\end{align}
\blue{
and 
$[\phi_0(t, x, z), \phi_0(t',x', z')]$ is proportional to the retarded Green function for $t>t'$, which does not depend on the state. 

In the planar black hole (finite temperature) case, we will consider the expectation value at temperature $\beta$:
\begin{align}
     \langle \phi(t,x,z) \rangle_{\beta,\mathrm{wp}}= \Tr \left( e^{- \beta H} \phi_{p, \omega}^{\dagger} \phi(t, x, z) \phi_{p, \omega} \right),
\end{align}
where
\begin{align}
     \phi_{p, \omega} 
    &= \lim_{z \to 0} \frac{1}{z^\Delta} 
    \int dt \, d x \, e^{-\frac{t^2+x^2}{2a^2}- i \omega t+ip x} \phi(t, x, z).
\end{align}
We can evaluate this expectation value following the zero temperature case, and the result is the same as \eqref{vev} just replacing $D^+$ with the ones in the planar black hole case.}
There could be a choice for the vacuum for the planar black hole background, but we need to choose the Hartle-Hawking vacuum because our computation in the CFT is based on the analytic continuation from the Euclidean theory. Then, 
\re{we need to use the following formula~\cite{Louko:2000tp,Parikh:2012kg}:}
\re{
\begin{align}
    D^+_{\rm BH}(X,X')
    &\sim
    \frac{1}{2\pi L}
    \left[
        \frac{zz'}
        {4z_h^2\,
        \sinh\!\left(\frac{(x-x')+(t-t'-i\epsilon)}{2z_h}\right)
        \sinh\!\left(\frac{(x-x')-(t-t'-i\epsilon)}{2z_h}\right)}
    \right]^\Delta,
    \qquad z,z'\to0,
    \label{bhret}
\end{align}
where $2 \pi z_h= \beta$.
}
\blue{if both $X,X'$ are on the asymptotic boundary.
From \eqref{bhret}, we find that the bulk computation yields exponential thermal damping.
}

\blue{Finally, we will evaluate the integrals in  \eqref{vev} approximately and show the consistency with  CFT results. Below, we consider the zero-temperature case; it is straightforward to generalize this to finite temperature by replacing the propagator with the thermal one.}
The overlap 
\begin{align}
     \lim_{z_1 \to 0} \frac{1}{z_1^\Delta} 
    \int dt_1 \, d x_1 \, e^{-\frac{t_1^2+x_1^2}{2a^2} + i \omega t_1 - ip x_1} D^+(X_1,X'),
    \label{ov}
\end{align}
is proportional to $e^{-\frac{a^2}{2} ( (\delta k)^2+\delta \omega^2)}$ as shown in \eqref{ov3}, then, it is localized on the wave packet trajectory by the Gaussian factor.
Thus, the first term in \eqref{vev} gives the contribution by the retarded propagator from the source localized in the wave packet trajectory, which may be the shockwave-like contribution.
In particular, this (almost) vanishes for $|x| \gg a$ at $t=z=0$.
The second and third terms in \eqref{vev} also include the overlap \eqref{ov}, and the $X'$-integrations are localized near the trajectory. The retarded propagator in those terms will be nonzero for $t'=x'=z'=0$ because $t_1$ or $t_3$ is localized near $t_1=0$ or $t_3=0$, and then the factor $D^+(X,X')$ can be approximated as the bulk-boundary propagator. In particular, 
$D^+(X,X') \simeq 1/(uv)^\Delta$ at $z=0$.
These are consistent with the above general discussions \blue{and the CFT results}.

\section*{Acknowledgement}

The authors would like to thank Daichi Takeda for the useful comments.
This work was supported by MEXT-JSPS Grant-in-Aid for Transformative Research Areas (A) ``Extreme Universe'', No.\ 21H05184,
and
JSPS KAKENHI Grant Number 24K07048.
The work of N.~T.~was supported in part by JSPS KAKENHI Grant No.~JP21H05189, JP22H05111 and 25K07282.

\appendix

\section{Localized excited state} \label{shws}

Here, we consider the localized excited state introduced in~\cite{Caputa:2014eta,Nozaki:2014hna}, which is a local primary-operator insertion with a small Euclidean-time evolution.
We will compute the expectation value of the primary operator for the state, which corresponds to a bulk-propagating shockwave, as discussed in~\cite{Afkhami-Jeddi:2017rmx}, instead of a wave packet, both at zero and finite temperature. 
To construct the CFT state dual to a spherical shockwave propagating from the boundary into the bulk, we consider a pure state created by the insertion of a heavy local operator:
\begin{align}
    \ket{\psi_0} = \cO(x_0)\ket{0}, \quad \bra{\psi_0} = \bra{0} \cO(x_0^*),
\end{align}
where \(\Delta\) is taken to be large so that the state \(\ket{\psi_0}\) corresponds to a shockwave, although the following computations will be correct without assuming its largeness. Since an insertion \(\cO(x)\) at a Lorentzian spacetime point leads to a non-normalizable state, we instead insert the operator at a Euclidean time \(t'=i\delta\), which corresponds to the point
\begin{align}
x_0 = (u_0', v_0') = (i\delta, i\delta),\quad x_0^* = (-i\delta, -i\delta)
\end{align}
in null coordinates \(u' = t' + x'\) and \(v' = t' -x'\) with the metric \(ds^2 = -du'dv'\).

Using this state, we can easily calculate the expectation value of \(\cO(x)\):
\begin{align}
        \langle \cO(x) \rangle_{0,\mathrm{sw}} & = \frac{\bra{\psi_0}\cO(x)\ket{\psi_0}}{\bra{\psi_0}\psi_0\rangle} \nn
        & = \frac{\lr{-1}^\frac{3\Delta}{2}\lr{2\delta}^\Delta C_{123}}{\lr{u^2+\delta^2}^{\frac{\Delta}{2}}\lr{v^2+\delta^2}^{\frac{\Delta}{2}}}.
\end{align}
As in the case of the wave packet, this result indicates a power-law suppression of the expectation value along the light cone.\\
We extend this analysis to the finite-temperature case.
\begin{align}
    \langle \cO(x^*_0) \cO(x_0) \rangle_\beta & = \lr{\frac{\pi}{\beta}}^{2\Delta} \frac{1}{\sinh^\Delta \frac{\pi(-2i\delta)}{\beta}\sinh^\Delta\frac{\pi(2i\delta)}{\beta}}, \nn
    \langle \cO(x^*_0) \cO(x) \cO(x_0) \rangle_\beta & = C_{123}\lr{\frac{\pi}{\beta}}^{3\Delta} \frac{1}{ \sinh^{\frac{\Delta}{2}} \left( \frac{\pi (-i\delta - u)}{\beta} \right)} \frac{1}{ \sinh^{\frac{\Delta}{2}} \left( \frac{\pi (u - i\delta)}{\beta} \right)} \frac{1}{ \sinh^{\frac{\Delta}{2}} \left( \frac{\pi (2i\delta)}{\beta} \right)} \nn
        &\quad \times
        \frac{1}{ \sinh^{\frac{\Delta}{2}} \left( \frac{\pi (v + i\delta)}{\beta} \right)} 
        \frac{1}{ \sinh^{\frac{\Delta}{2}} \left( \frac{\pi (i\delta - v)}{\beta} \right)}\frac{1}{ \sinh^{\frac{\Delta}{2}} \left( \frac{\pi (-2i\delta)}{\beta} \right)}
\end{align}
As a result,
\begin{align}
        \langle \cO(x) \rangle_{\beta,\mathrm{sw}} &= \frac{\langle \cO(x^*_0) \cO(x) \cO(x_0) \rangle_\beta}{\langle \cO(x^*_0) \cO(x_0) \rangle_\beta} \nn
     & =(-1)^\frac{3\Delta}{2} C_{123} \lr{\frac{\pi}{\beta}}^\Delta \frac{\sin^\Delta\frac{2\pi\delta}{\beta}}{\lr{\sinh^2 \frac{\pi u}{\beta}+\sin^2\frac{\pi \delta}{\beta}}^\frac{\Delta}{2}\lr{\sinh^2 \frac{\pi v}{\beta}+\sin^2\frac{\pi \delta}{\beta}}^\frac{\Delta}{2}}.
     \label{ols}
\end{align}
This result is similar to the wave packet analysis and exhibits temperature-dependent exponential decay along the light cone.
For the wave packet case, there are three different kinds of contributions, which are localized on $u=v=0$, $u=0$, or $v=0$ and spread inside the light cone.
The result here \eqref{ols} can be separated for $u={\cal O}(\delta),v={\cal O}(\delta) $ region, and $u={\cal O}(\delta)$ or $v={\cal O}(\delta) $ regions, and $uv \gg \delta^2$ region.
These are analogous terms in \eqref{evo}, although there is no $1/(pa)$ hierarchy because the momentum $p$ could be regarded as the UV cut-off $1/\delta$ here.

\paragraph*{A bulk picture}

Here, we compute the overlap between this state and a bulk local excitation to understand how this behaves in the bulk picture.
We work in Poincar\'e AdS$_3$ with metric
\begin{equation}
  ds^2 \;=\; \frac{\ell^2}{z^2}\left(-dt^2 + dx^2 + dz^2\right),
\end{equation}
where the conformal boundary is located at $z=0$. A scalar bulk field $\Phi$
of mass $m$ is dual to a scalar primary operator $\mathcal{O}$ in the
boundary CFT with scaling dimension
\begin{equation}
  \Delta(\Delta-2) \;=\; m^2 \ell^2.
\end{equation}
We consider the state created by acting with the CFT operator at the origin,
\begin{equation}
  \ket{\psi_0} =e^{-H \delta}  \mathcal{O}(x_0)\,\ket{0},
\end{equation}
with the Euclidean time smearing, 
where $H$ is the CFT Hamiltonian. We probe this state with a bulk local excitation at the point
\begin{equation}
  Y = (z, t_Y=0, x_Y),
\end{equation}
by computing the amplitude
\begin{equation}
  A(t;z,x_Y) \;\equiv\; \bra{0} \Phi(z,0,x_Y)\,e^{iH(t+i\delta)}\mathcal{O}(0,0) \ket{0}.
\end{equation}
This amplitude can be written as a mixed bulk--boundary two-point
function:
\begin{equation}
  A(t;z,x_Y)
  \;=\;
  \bra{0} \Phi(z,0,x_Y)\,\mathcal{O}(t+i\delta,0) \ket{0}.
\end{equation}

In the free bulk field limit, this correlator is given by the
bulk--to--boundary Wightman propagator in Poincar\'e AdS$_3$. For a scalar of
dimension $\Delta$ one has
\begin{equation}
  \bra{0}\,\Phi(z,t,x)\,\mathcal{O}(t_b,x_b)\,\ket{0}
  \;\propto\;
  \left[
    \frac{z}{
      z^2 - (t - t_b - i\epsilon)^2 + (x - x_b)^2
    }
  \right]^{\Delta},
\end{equation}
where the $i\epsilon$ prescription specifies the Wightman (rather than
time-ordered) correlator.

Specializing to our choice of insertion points,
\begin{equation}
  t = 0,\quad x = x_Y,\quad t_b = t+i\delta,\quad x_b = 0,
\end{equation}
we obtain
\begin{equation}
  A(t;z,x_Y)
  \;\propto\;
  \left[
    \frac{z}{
      z^2 + x_Y^2 - (t + i\delta)^2
    }
  \right]^{\Delta}.
\end{equation}

Thus the overlap between the bulk local excitation $\Phi(z,0,x_Y)\ket{0}$ and
the time-evolved boundary excitation $e^{iH(t+i\epsilon)}\mathcal{O}(0,0)\ket{0}$
is precisely the bulk--boundary two-point function evaluated at a complex time
separation $t+i\epsilon$. Physically, $A(t;z,x_Y)$ plays the role of the
``wavefunction in the bulk'' of the state created by $\mathcal{O}(0,0)$ and
propagated for a complex time $t+i\epsilon$.

We can see that $A$ is large near the light cone $z^2+x_Y^2-t^2=0$.
In particular, for $\Delta \gg 1$, it is almost localized on the light cone.
Near the light cone, with $\rho=\sqrt{z^2+x_Y^2}=t+\delta \rho$, 
we have $A \sim \left(\frac{z}{2 t (\delta \rho - i \delta)} \right)^\Delta$.
The region where $|A|$ is large 
is $\delta \rho \sim \delta$, and $|A|$ at $\delta \rho =0$ is proportional to $(z/t)^\Delta$, which is order 1
for $1-z/t \sim 1/\Delta$.
Thus, this state represents a localized excitation over the light cone within a very small angle proportional to
\NTmod{$1/\Delta$,}
which is consistent with the one in~\cite{Caputa:2014eta,Nozaki:2014hna,Nozaki:2013wia},
where this state is supposed to represent null geodesics on $x=0$ in the bulk picture.

\section{Zero-temperature three-point functions for generic \(\Delta\)} \label{gendelt}

\NTmod{We evaluate the expectation value of a primary operator $\mathcal{O}(t,x)$ with respect to the wave-packet state at zero temperature for generic conformal dimension $\Delta$ in this appendix. This supplements the analysis in section~\ref{sec:zero-temperature-cft}, in which the expectation value for $\Delta=2$ is derived.}

First, the normalization constant for generic \(\Delta\) at zero temperature is calculated as
\begin{align}
    {\cal N}^2_0 = &\bra{p,\bar\omega}   p,\bar\omega \rangle \nn
   =
   & \int d t_1 \, d x_1 \, e^{-\frac{ (x_1)^2+t_1^2}{2 a^2}-i p x_1+i \bar{\omega} t_1} 
   \int d t_2 \, d x_2 \, e^{-\frac{ (x_2)^2+t_2^2}{2 a^2}+i p x_2-i \bar{\omega} t_2} 
    \frac{1}{(u_1-u_2 \NTmod{- i\epsilon})^{ \Delta} (v_2-v_1 \NTmod{+ i\epsilon})^{ \Delta} } \nn
    =
    & \frac{1}{4}\int d u_1 \, d v_1 \, d u_2 \, d v_2 \,\, e^{-\frac{ (u_1-i p_u a^2)^2+(v_1-i p_v a^2)^2+(u_2+i p_u a^2)^2+(v_2+i p_v a^2)^2}{4 a^2} 
    - \frac{a^2}{2} ((p_u)^2+(p_v)^2) }
     \nn
   & \times 
    \frac{1}{
    (u_1-u_2 \NTmod{- i\epsilon})^{ \Delta} (v_2-v_1 \NTmod{+ i\epsilon})^{ \Delta}} \nn
    \simeq
    & \frac{1}{4} \,\,
     \frac{(2 \pi)^{3/2}  }{\Gamma(\Delta)} (-i)^{\Delta}
      a (p_v/2)^{\Delta-1} \times 
      \frac{(2 \pi)^{3/2}  }{\Gamma(\Delta)} (i)^{\Delta}
      a (p_u/2)^{\Delta-1}.
\end{align}
Then, we proceed to calculate \(\bra{p,\omega}\cO\ket{p,\omega}\):
\begin{align}
     & \bra{p,\omega}   \cO(t=t_2, x=x_2) \ket{p,\omega} \nn
   =
   & \frac{1}{4}\int d u_1 \, d v_1 \, d u_3 \, d v_3 \,\, e^{-\frac{ (u_1-i a^2p_u)^2+(v_1-i a^2p_v)^2+(u_3+i  a^2p_u)^2+(v_3+i a^2p_v)^2}{4 a^2} 
     - \frac{a^2}{2} ((p_u)^2+(p_v)^2) }   \nn
   &\times C_{123}\frac{1}{u_{12}^{\Delta/2} u_{23}^{\Delta/2} u_{31}^{\Delta/2}} \frac{1}{v_{21}^{\Delta/2} v_{32}^{\Delta/2} v_{13}^{\Delta/2}} .
\end{align}
\NTmod{The $u$-part of this integral is evaluated as follows.}
\md{A direct application of the residue theorem is cumbersome, as it requires considering derivatives acting on both the exponential function and the 
\NTmod{polynomials}
in the denominator.
\NTmod{
We avoid this complication by 
expressing the integrand as a derivative of a function with simple poles with respect to $\alpha\equiv -i\ep_{12}$ and $\gamma\equiv -i\ep_{13}$. We then apply the residue theorem, which significantly simplifies the calculation.
With this approach, we find}
}
\begin{align}
    &(-1)^{-\Delta/2}\int d u_1  \, d u_3  \,\, e^{-\frac{ (u_1-i a^2p_u)^2+(u_3+i  a^2p_u)^2}{4 a^2} 
    - \frac{a^2p_u^2}{2}  } 
    \frac{1}{u_{12}^{\Delta/2} u_{23}^{\Delta/2} u_{13}^{\Delta/2}} \nn
   & \simeq (-1)^{-\Delta/2}\int du_3 \Bigg\{
   \NTmod{\frac{2\pi i}{\Gamma(\frac\Delta2)}}
   \frac{\partial^{\Delta/2-1}}{\partial \alpha^{\Delta/2-1}}
   \frac{e^{-\frac{\lr{u_2-ia^2p_u}^2}{4a^2}}}{u_{23}^\Delta}
   \NTmod{ e^{\frac{i\alpha p_u}{2}} }
   e^{-\frac{\lr{u_3+ia^2p_u}^2}{4a^2}}
   e^{-\frac{a^2p_u^2}{2}}
   \nn
   & \hphantom{ \simeq(-1)^{-\Delta/2}\int du_3 }
   +
   \NTmod{\frac{2\pi i}{\Gamma(\frac\Delta2)}}
   \frac{\partial^{\Delta/2-1}}{\partial \gamma^{\Delta/2-1}}(-1)^{\Delta/2}\frac{e^{-\frac{(u_3)^2}{2a^2}}}{u_{32}^\Delta}
   \NTmod{e^{\frac{i\gamma p_u }{2}}}
   \Bigg\} \nn
   &\simeq
   \NTmod{\frac{4\pi^2}{\Gamma(\frac\Delta2)\Gamma(\Delta)} \lr{\frac{ip_u}{2}}^{\Delta/2-1}
   (-1)^{-\frac{3}{2}\Delta }
   }
   \lr{\frac{ip_u}{2}}^{\Delta-1}e^{-\frac{u_2^2}{2a^2}}
   +
   \NTmod{\frac{2\pi i}{\Gamma(\frac\Delta2)}\lr{\frac{ip_u}{2}}^{\Delta/2-1}(-1)^{\Delta}}
   \frac{a\sqrt{2\pi}}{u_2^\Delta},
\end{align}
where the second term of the \(u_3\) integrals is obtained using the saddle-point approximation and is valid only for \(|u| \gg a\).
The integrals over \(v_1\) and \(v_3\) can be performed in the same way, up to an overall sign.
%
\NTmod{Combining with the $v$-part of the integral, which is evaluated in a similar manner,}
we obtain
\begin{align}
  &  \langle \cO(t,x)\rangle_{0,\mathrm{wp}}  = \frac{\bra{p,\omega}\cO(t,x)\ket{p,\omega}}{\cN_0^2} \nn
    %
    &\simeq
    \NTmod{
    \frac{C_{123}}{\Gamma(\frac\Delta2)^2}
    \Biggl\{
    (-1)^{-\Delta} \frac{2\pi}{a^2} \lr{\frac{p_up_v}{4}}^{\Delta/2-1}e^{-\frac{u^2+v^2}{2a^2}}
    }
    \nn
    &
    \NTmod{
    \hphantom{ \frac{C_{123}}{\Gamma(\frac\Delta2)^2}\Biggl\{}
    + \frac{2\sqrt{2\pi}}{a}(-1)^{-2\Delta}\Gamma(\Delta)
    \left(
        \frac{p_u^{\Delta/2-1}}{p_v^{\Delta/2}}
        \frac{ e^{-\frac{u^2}{2a^2}} }{v^\Delta }
        + \frac{p_v^{\Delta/2-1}}{p_u^{\Delta/2}}
        \frac{ e^{-\frac{v^2}{2a^2}} }{u^\Delta }
    \right)
    +\frac{ (-1)^{-3\Delta}\Gamma(\Delta)^2}{\lr{\frac{p_up_v}{4}}^{\Delta/2}u^\Delta v^\Delta}
    \Biggr\}.
    }
    \label{generalD}
\end{align}

\section{Finite-temperature three-point function for \(\zeta \gg1\)} \label{zetagg}
We here provide the details of the calculation of the finite-temperature three-point function for \(\zeta \gg 1\) that were omitted in section~\ref{ftcft}.

The \(u_3\)-integration now becomes
\begin{align}
    F_3 &\coloneqq \int du_3 \, F_1 e^{-\frac{(u_3+ip_ua^2)^2}{4a^2}
        -\frac{a^2p_u^2}{2}} \frac{1}{\sinh \frac{\pi(u-u_3-i \epsilon)}{\beta}} \nn
    & = \int du_3 \, e^{-\frac{(u_3+ip_ua^2)^2}{4a^2}
        -\frac{a^2p_u^2}{2}}(-1) \nn
        & \times \Bigg\{(2\pi i) \sum_{n=0}^{\lf{\zeta}}\frac{\beta}{\pi}
        \frac{e^{-\frac{(u+i\beta(n-\zeta))^2}{4a^2}}}{\sinh^2 \left( \frac{\pi (u_3 - u+i\epsilon)}{\beta} \right)} 
        + (2\pi i) \sum_{n=0}^{\lf{\zeta}}
       \lr{-\frac{\beta}{\pi}}\frac{e^{-\frac{(u_3+i\beta(n-\zeta))^2}{4a^2}}}{\sinh^2 \left( \frac{\pi (u_3 - u+ i \epsilon)}{\beta} \right)} \Bigg\} \nn
    & \simeq 4 \pi^2 
        \sum_{l=0}^{\lf{\zeta}} \sum_{n=0}^{\lf{\zeta}}\lr{\frac{\beta}{\pi}}^3
        \lr{-\frac{1}{2a^2}i\beta(l-\zeta)}e^{-\frac{1}{2a^2}\lr{u-\frac{i\beta}{2}(l-n)}^2}e^{-\frac{\beta^2(4\zeta-l-n)(l+n)}{8a^2}}\nn
        & + 2\pi i \,\frac{\beta}{\pi}\sqrt{2\pi}a \sum_{n=0}^{\lf{\zeta}}\frac{e^{-\frac{\beta^2n(4\zeta-n)}{8a^2}}}{\sinh^2\lr{\frac{\pi\lr{\frac{i\beta n}{2}-u}}{\beta}}} . \label{F3}
\end{align}
The first term was obtained by the residue theorem, while the second term was obtained via the saddle-point approximation. Note that this approximation is valid in the regime \(|u|\gg a\). We approximate the sum as follows:
\begin{align}
     &\sum_{l=0}^{\lf{\zeta}} \sum_{n=0}^{\lf{\zeta}}-\frac{1}{2a^2}\lr{i\beta(l-\zeta)}e^{-\frac{1}{2a^2}\lr{u-\frac{i\beta}{2}(l-n)}^2}e^{-\frac{\beta^2(4\zeta-l-n)(l+n)}{8a^2}} \nn
    = & \sum_{l=0}^{\lf{\zeta}} \sum_{n=0}^{\lf{\zeta}}-\frac{1}{2a^2}\lr{i\beta(l-\zeta)}e^{-\frac{u^2}{2a^2}}e^{-\frac{1}{4a^2}\lr{2\lr{\beta^2\zeta+i\beta u}n-\beta^2n^2}}e^{-\frac{1}{4a^2}\lr{2\lr{\beta^2\zeta-i\beta u}l-\beta^2l^2}} \nn
    \simeq & \sum_{l=0}^{\lf{\zeta}}-\frac{1}{2a^2}\lr{i\beta(l-\zeta)}e^{-\frac{u^2}{2a^2}}e^{-\frac{1}{4a^2}\lr{2\lr{\beta^2\zeta-i\beta u}l-\beta^2l^2}}\frac{1}{1-e^{-\frac{\beta p_u}{2}}}\nn
    \simeq & e^{-\frac{u^2}{2a^2}}\frac{i\beta}{2a^2}\lr{\frac{\zeta}{1-e^{-\frac{\beta p_u}{2}}}-\frac{e^{-\frac{\beta p_u}{2}}}{\lr{1-e^{-\frac{\beta p_u}{2}}}^2}}\frac{1}{1-e^{-\frac{\beta p_u}{2}}}\nn 
    \simeq & e^{-\frac{u^2}{2a^2}}\frac{ip_u}{2}\lr{\frac{1}{1-e^{-\frac{\beta p_u}{2}}}}^2.
\end{align}
In the third equality, we neglected the quadratic term \MD{of \(n\) in} the summation, since \(2\beta^2\zeta n \gg \beta^2n^2\), and treated it as a geometric series. Note that we assume \(\zeta \gg 1\). 
\MD{Since \(\zeta\) roughly represents the number of contributing poles that appear due to finite-temperature effects, this assumption implies the high-temperature limit.}
In the fourth equality, we used the relation \(\sum_{l} l e^{-\frac{\beta p_u}{2}l}=-2\frac{\partial}{\partial(\beta p_u)}\sum_l e^{-\frac{\beta p_u}{2}l}\). In the fifth equality, we neglect the second term, since we assume \(\zeta \gg1\).\\
We now perform the summation in the second term of \eqref{F3}.
\begin{align}
    \sum_{n=0}^{\lf{\zeta}}\frac{e^{-\frac{\beta^2n(4\zeta-n)}{8a^2}}}{\sinh^2\lr{\frac{\pi\lr{\frac{i\beta n}{2}-u}}{\beta}}}
    &\simeq  \frac{\frac{1-(e^{-\beta p_u})^{\lf{\zeta/2}}}{1-e^{-\beta p_u}}}{\sinh^2\frac{\pi u}{\beta}} - \frac{\frac{1-(e^{-\beta p_u})^{\lf{\frac{\zeta-1}{2}}}}{1-e^{-\beta p_u}}}{\cosh^2\frac{\pi u}{\beta}}e^{-\frac{\beta p_u}{2}}e^{\frac{\beta^2}{8a^2}} \nn
    & \simeq  \frac{\frac{1}{1-e^{-\beta p_u}}}{\sinh^2\frac{\pi u}{\beta}} - \frac{\frac{1}{1-e^{-\beta p_u}}}{\cosh^2\frac{\pi u}{\beta}}e^{-\frac{\beta p_u}{2}}
    \NTmod{e^{\frac{\beta^2}{8a^2}} }  ,
\end{align}
where the summation is separated into even and odd parts. To extract the leading contribution, it is sufficient to consider the \(n=0\) term.

Apart from the overall sign, the integration over \(v\) proceeds in exactly the same way, and the result is summarized as follows:
\begin{align}
    &\langle\cO(t,x) \rangle_{\beta,\mathrm{wp}}\nn
    = & 4C_{123}\pi^4\lrm{e^{-\frac{u^2}{2a^2}}\frac{ip_u}{2}\lr{\frac{1}{1-e^{-\frac{\beta p_u}{2}}}}^2
    +\frac{i\pi  a}{\beta^2}\sqrt{\frac{\pi}{2}}\lr{\frac{\frac{1}{1-e^{-\beta p_u}}}{\sinh^2\frac{\pi u}{\beta}} - \frac{\frac{1}{1-e^{-\beta p_u}}}{\cosh^2\frac{\pi u}{\beta}}e^{-\frac{\beta p_u}{2}}e^{+\frac{\beta^2}{8a^2}}}}\nn
    \times&\lrm{-e^{-\frac{v^2}{2a^2}}\frac{ip_v}{2}\lr{\frac{1}{1-e^{-\frac{\beta p_v}{2}}}}^2
    -\frac{i\pi  a}{\beta^2}\sqrt{\frac{\pi}{2}}\lr{\frac{\frac{1}{1-e^{-\beta p_v}}}{\sinh^2\frac{\pi v}{\beta}} - \frac{\frac{1}{1-e^{-\beta p_v}}}{\cosh^2\frac{\pi v}{\beta}}e^{-\frac{\beta p_v}{2}}e^{+\frac{\beta^2}{8a^2}}}}/\mathcal{N}_\beta^2, \nn
    \label{obeta}
\end{align}
where \(\cN_\beta^2\) is defined as
\begin{align}
        \mathcal{N}^2_\beta \coloneqq& \frac{1}{4}\int du_1 dv_1 du_3 dv_3 e^{-\frac{(u_1-ip_u a^2)^2 + (v_1-ip_va^2)^2+(u_3+ip_ua^2)^2+(v_3+ip_va^2)^2}{4a^2}
    -\frac{a^2p_u^2}{2}-\frac{a^2p_v^2}{2}} \nn
    &\times \left( \frac{\pi}{\beta}\right)^{2\Delta} 
            \frac{1}{ \sinh^\Delta \frac{\pi(u_1-u_3-i\epsilon)}{\beta}\sinh^\Delta \frac{\pi(v_3 -v_1 + i\epsilon)}{\beta}} \nn \simeq &\frac{\pi^3 a^2}{2^{2\Delta-3}}\frac{(p_u p_v)^{\Delta-1}}{\Gamma(\Delta)^2}\sum_{m=0}^{\lfloor \zeta \rfloor}\sum_{n=0}^{\lfloor  \xi \rfloor}
    (-1)^{-(n+m)\Delta}e^{-\frac{\beta^2 n(4\xi- n)}{8a^2}}e^{-\frac{\beta^2 m(4\zeta- m)}{8a^2} }\nn
     \simeq & \frac{\pi^3 a^2}{2} \, p_u p_v  \sum_{m=0}^{\lfloor \zeta \rfloor}\sum_{n=0}^{\lfloor  \xi \rfloor}e^{-\frac{\beta^2 n(4\xi- n)}{8a^2}}e^{-\frac{\beta^2 m(4\zeta- m)}{8a^2} }\nn 
     \simeq& \frac{\pi^3 a^2}{2}\, p_u p_v \,\frac{1}{1-e^{-\frac{\beta p_u}{2}}}\frac{1}{1-e^{-\frac{\beta p_v}{2}}}.
     \label{OfiniteT}
\end{align}
We arrive at the following expression:
\begin{align}
    &\langle\cO(t,x) \rangle_{\beta,\mathrm{wp}}\nn
    \simeq& 2C_{123} \Bigg(
    e^{-\frac{u^2}{2a^2}}e^{-\frac{v^2}{2a^2}}\frac{\pi}{a^2}\frac{1}{1-e^{-\frac{\beta p_u}{2}}}\frac{1}{1-e^{-\frac{\beta p_v}{2}}}
    + e^{-\frac{u^2}{2a^2}}\frac{\pi^2}{ap_v\beta^2}\sqrt{2\pi}\frac{1}{\sinh^2\frac{\pi v}{\beta}}\frac{1}{1-e^{-\frac{\beta p_u}{2}}}\nn
    &
    \hphantom{2C_{123} }
    + e^{-\frac{v^2}{2a^2}}\frac{\pi^2}{ap_u\beta^2}\sqrt{2\pi}\frac{1}{\sinh^2\frac{\pi u}{\beta}}\frac{1}{1-e^{-\frac{\beta p_v}{2}}}
    +\frac{2\pi^4}{\beta^4 p_up_v}\frac{1}{\sinh^2\frac{\pi u}{\beta}\sinh^2\frac{\pi v}{\beta}}
    \Bigg).\label{expotg_App}
\end{align}

\section{Details of calculation of energy density for \(\Delta = 3\)}\label{ded}

\NTmod{In this appendix, we show the derivation of the energy density of the wave packet state at finite temperature with the conformal dimension $\Delta=3$, which reproduces the expressions of the energy density; \eqref{energy-density_finite-T} at general temperature and \eqref{calE_AdS3CFT2} with \eqref{energy-density_zeroT-part} for zero temperature.}

\NTmod{We start our calculation from \eqref{A_beta-def}, which corresponds to a part of the expectation value of $T(u)$ (the first term on the right-hand side of \eqref{OTO_beta}).}
\begin{align}
&\langle A\rangle_{\beta,\mathrm{wp}}\nn
    \coloneqq&  \frac{1}{4}\int du_1 dv_1 du_3 dv_3 e^{-\frac{(u_1-ip_u a^2)^2 + (v_1-ip_va^2)^2+(u_3+ip_ua^2)^2+(v_3+ip_va^2)^2}{4a^2}
    -\frac{a^2p_u^2}{2}-\frac{a^2p_v^2}{2}}A_\beta\nn
    =&  \frac{1}{4}\int du_1 dv_1 du_3 dv_3 e^{-\frac{(u_1-ip_u a^2)^2 + (v_1-ip_va^2)^2+(u_3+ip_ua^2)^2+(v_3+ip_va^2)^2}{4a^2}
    -\frac{a^2p_u^2}{2}-\frac{a^2p_v^2}{2}} \nn
    & \times  \left( \frac{\pi}{\beta }\right)^2  \left( \frac{\pi}{\beta}\right)^{2\Delta} 
       \frac{\Delta}{2}
    \frac{\sinh^2 \frac{\pi(u_1-u_3-i\ep)}{\beta}}{\sinh^2 \frac{\pi(u-u_1+i\ep)}{\beta}\sinh^2 \frac{\pi(u-u_3-i\ep)}{\beta}}
    \frac{1}{ \sinh^\Delta \frac{\pi(u_1-u_3-i\ep)}{\beta}\sinh^\Delta \frac{\pi(v_3-v_1+i\ep)}{\beta}}.
\end{align}
\(v_1\) and \(v_3\)-integrations are performed as
\begin{equation}
    \int dv_1 dv_3 \frac{1}{\sinh^3\frac{\pi(v_3-v_1+i\ep)}{\beta}}e^{-\frac{(v_1-ip_va^2)^2+(v_3+ip_va^2)^2}{4a^2}
    -\frac{a^2p_v^2}{2}} 
    \simeq
    \lr{-2\pi i}a\sqrt{2\pi}\lr{\frac{\beta}{\pi}}^3\lr{-\frac{p_v^2}{\MD{8}}}\frac{1}{
    \NTmod{1+e^{-\frac{\beta p_v}{2}}}
    } .
\end{equation}
Next, we consider the \(u_1\) and \(u_3\) integrations.
\begin{align}
    &\int du_1 du_3  e^{-\frac{(u_1-ip_u a^2)^2 +(u_3+ip_ua^2)^2}{4a^2}
    -\frac{a^2p_u^2}{2}} \frac{1}{\sinh^2\frac{\pi\lr{u-u_1+i\ep}}{\beta}\sinh^2\frac{\pi\lr{u-u_3-i\ep}}{\beta}\sinh\frac{\pi\lr{u_1-u_3-i\ep}}{\beta}} \nn
    \simeq& \int du_3 e^{-\frac{(u_3+i\beta\zeta)^2}{4a^2}-\frac{\beta^2\zeta^2}{2a^2}}\frac{1}{\sinh^2\frac{\pi\lr{u-u_3}}{\beta}} \nn
    &\times 2\pi i \Biggl\{
    \lr{\frac{\beta}{\pi}}^2\sum_{n=0}^{\lf{\zeta}}
    \biggl[
    -\frac{1}{2a^2}\lr{i\beta n-i\beta\zeta}(-1)^n\frac{e^{-\frac{(u+i\beta n-i\beta \zeta)^2}{4a^2}}}{\sinh\frac{\pi\lr{u-u_3}}{\beta}}
    \nn
    & \hphantom{\times 2\pi i}
    +e^{-\frac{(u+i\beta n -i\beta \zeta)^2}{4a^2}} (-1)^{n+1}\frac{\cosh\frac{\pi\lr{u-u_3}}{\beta}}{\sinh^2\frac{\pi\lr{u-u_3}}{\beta}}\frac{\pi}{\beta}
    \biggr]
     + \frac{\beta}{\pi}\sum_{n=0}^{\lf{\zeta}}
     \NTmod{(-1)^n}
     \frac{e^{-\frac{\lr{u_3+i\beta n -i\beta \zeta}^2}{4a^2}}}{\sinh^2\frac{\pi\lr{u_3-u+i\beta n}}{\beta} }     \Biggr\}
\end{align}
The first term in the curly brackets can be evaluated as
\begin{align}   
    &\sum_{n=0}^{\lf{\zeta}}(-1)^{n+1}(2\pi i)\lr{\frac{\beta}{\pi}}^2e^{-\frac{\beta^2\zeta^2}{2a^2}}e^{-\frac{(u+i\beta n -i\beta\zeta)^2}{4a^2}}\lr{-\frac{i\beta n -i\beta \zeta}{2a^2}}\int du_3 
    \frac{e^{-\frac{\lr{u_3+i\beta\zeta}^2}{4a^2}}}{\sinh^3\frac{\pi\lr{u_3-u}}{\beta}}  \nn
   \simeq &\sum_{n=0}^{\lf{\zeta}} (-1)^{n+1}(4\pi^2)\lr{\frac{\beta}{\pi}}^2e^{-\frac{\beta^2\zeta^2}{2a^2}}e^{-\frac{(u+i\beta n -i\beta\zeta)^2}{4a^2}}\lr{-\frac{i\beta n -i\beta \zeta}{2a^2}}
   \nn & \times 
   \frac{1}{2}\sum_{l=0}^{\lf{\zeta}}
   \NTmod{(-1)^l}
   \lr{\frac{\beta}{\pi}}^3\lr{-\frac{-i\beta l +i\beta\zeta}{2a^2}}^2e^{-\frac{\lr{u-i\beta l +i\beta\zeta}^2}{4a^2}}\nn
   \simeq & e^{-\frac{u^2}{2a^2}} \sum_{n=0}^{\lf{\zeta}}\sum_{l=0}^{\lf{\zeta}}
   \NTmod{(-1)^{n+l+1}}
   (2\pi^2)\lr{\frac{\beta}{\pi}}^5e^{-\frac{\beta p_u}{2}n}e^{-\frac{\beta p_u}{2}l} \lr{-\frac{i\beta n-i\beta \zeta}{2a^2}}\lr{-\frac{-i\beta l+i\beta \zeta}{2a^2}}^2 \nn
   \simeq & e^{-\frac{u^2}{2a^2}}(2\pi^2)\lr{\frac{\beta}{\pi}}^5
   \NTmod{
   \lr{\frac{p_u^2}{4}\frac{1}{1+e^{-\frac{\beta p_u}{2}}}+\frac{\beta p_u}{2a^2}\frac{e^{-\frac{\beta p_u}{2}}}{\lr{1+e^{-\frac{\beta p_u}{2}}}^2}}\lr{\frac{ip_u}{2}\frac{1}{1+e^{-\frac{\beta p_u}{2}}}+\frac{i\beta}{2a^2}\frac{e^{-\frac{\beta p_u}{2}}}{\lr{1+e^{-\frac{\beta p_u}{2}}}^2}}
   }
   \nn
   \simeq & 4\pi^2 i e^{-\frac{u^2}{2a^2}}    \lr{\frac{\beta}{\pi}}^5
   \NTmod{
    \lr{
        \frac{p_u^3}{16}\frac{1}{\left(1+e^{-\beta p_u/2}\right)^2}
        + \frac{3 p_u^2}{16}\frac{e^{-\beta p_u/2}}{\left(1+e^{-\beta p_u/2}\right)^2}
    }
   }
    + \cdots.
\end{align}
The second term can be evaluated as
\begin{align}
    &\sum_{n=0}^{\lf{\zeta}}(-1)^{n+1}(2\pi i)\frac{\beta}{\pi}e^{-\frac{\beta^2\zeta^2}{2a^2}}e^{-\frac{(u+i\beta n -i\beta\zeta)^2}{4a^2}} \int du_3 \frac{\cosh\frac{\pi(u_3-u)}{\beta}}{\sinh^4\frac{\pi\lr{u_3-u}}{\beta}}e^{-\frac{\lr{u_3+i\beta\zeta}^2}{4a^2}} \nn
    \simeq&\sum_{n=0}^{\lf{\zeta}}(-1)^{n+1}(4\pi^2)\frac{\beta}{\pi}e^{-\frac{\beta^2\zeta^2}{2a^2}}e^{-\frac{(u+i\beta n -i\beta\zeta)^2}{4a^2}}\sum_{l=0}^{\lf{\zeta}}\lr{\frac{\beta}{\pi}}^4\frac{1}{3!}
    \frac{\partial^3}{\partial u_3^3}
    \lr{\cosh\frac{\pi\lr{u_3-u}}{\beta}e^{-\frac{(u_3+i\beta\zeta)^2}{4a^2}}}\Bigg|_{u_3 =u-i\beta l} \nn
    =& \sum_{n=0}^{\lf{\zeta}} (-1)^{n+1}(4\pi^2)\lr{\frac{\beta}{\pi}}^5e^{-\frac{\beta^2\zeta^2}{2a^2}}e^{-\frac{(u+i\beta n -i\beta\zeta)^2}{4a^2}}\nn
    &\times\sum_{l=0}^{\lf{\zeta}}\frac{(-1)^l}{3!}\lrm{\lr{-\frac{-i\beta l +i\beta \zeta}{2a^2}}^3+3\lr{\frac{\pi}{\beta}}^2\lr{-\frac{-i\beta l+i\beta \zeta}{2a^2}}}e^{-\frac{\lr{u-i\beta l+i\beta \zeta}^2}{4a^2}} \nn
    \simeq & \sum_{n=0}^{\lf{\zeta}} (-1)^{n+1}(4\pi^2)\lr{\frac{\beta}{\pi}}^5e^{-\frac{u^2}{2a^2}}e^{-\frac{\beta^2\zeta}{2a^2}n} \nn
     &\times\sum_{l=0}^{\lf{\zeta}}\frac{(-1)^l}{3!}e^{-\frac{\beta^2\zeta}{2a^2}l}\lrm{\lr{-\frac{-i\beta l +i\beta \zeta}{2a^2}}^3+3\lr{\frac{\pi}{\beta}}^2\lr{-\frac{-i\beta l+i\beta \zeta}{2a^2}}} \nn
     \simeq & -4\pi^2 \lr{\frac{\beta}{\pi}}^5 e^{-\frac{u^2}{2a^2}}\frac{1}{1+e^{-\frac{\beta p_u}{2}}}i\lr{\frac{p_u^3}{48}\frac{1}{1+e^{-\frac{\beta p_u}{2}}}\NT{+}\frac{\beta p_u^2}{16a^2}\frac{e^{-\frac{\beta p_u}{2}}}{\lr{1+e^{-\frac{\beta p_u}{2}}}^2}}
\end{align}
The third term can be evaluated as:
\begin{equation}
    \sum_{n=0}^{\lf{\zeta}}(2\pi i)\frac{\beta}{\pi}
    \NTmod{(-1)^n}
    \int du_3\frac{e^{-\frac{1}{2a^2}\lr{u_3-\frac{i\beta n}{2}}^2}}{\sinh^4 \frac{\pi\lr{u_3-u+i\ep}}{\beta}} e^{-\frac{\beta^2}{8a^2}\lr{4\zeta-n}n} 
    \simeq 
    i2\sqrt{2\pi}a\beta \frac{1}{\sinh^4\frac{\pi u}{\beta}}+\cdots,
\end{equation}
where we use saddle-point approximation and only leave the leading contribution.\footnote{
A proper evaluation can be carried out following the same method used for the expectation value of 
\(\cO\), by decomposing the sum into even and odd components, neglecting the quadratic terms in \(l\), and applying a geometric series expansion.
}
The normalization constant for \(\Delta=3\) is given by the following expression:
\begin{align}
    \mathcal{N}_\beta^2\lr{\Delta=3} = \frac{\pi^3 a^2}{32}\lr{p_u p_v}^2
    \NTmod{\frac{1}{1+e^{-\frac{\beta p_u}{2}}}\frac{1}{1+e^{-\frac{\beta p_v}{2}}}.}
\end{align}
The expectation value of \(T(v)\) can be obtained through a nearly identical computation. Combining these results, we obtain
\begin{align}
    \mathcal{E}(t,x) &=\frac{\pi^2 c}{3\beta^2}
    \NTmod{
    +\frac{1}{2\sqrt{2\pi}a}
    \left(
        \frac{e^{-\frac{u^2}{2a^2}} p_u }{1+e^{-\beta p_u / 2}}
        + \frac{e^{-\frac{v^2}{2a^2}} p_v }{1+e^{-\beta p_v / 2}}
    \right)
    +
    \frac{6}{\pi}
    \lr{\frac{\pi}{\beta}}^4
    \left(
          \frac{1}{p_u^2\sinh^4\frac{\pi u}{\beta}}
        + \frac{1}{p_v^2\sinh^4\frac{\pi v}{\beta}}
    \right),
    }
\end{align}
where we have omitted the lower-order terms in the momentum expansion for
the 
second term, as the time dependence is contained entirely within \(e^{-\frac{u^2}{2a^2}}\) and  \(e^{-\frac{v^2}{2a^2}}\).

\section{Unitary wave packet operators} \label{unt}
At zero temperature, the expectation value of a primary field was given by:
\begin{equation}
    \langle \cO(t,x)\rangle_{0,\mathrm{wp}}
    \simeq
    2C_{123}\lr{\frac{\pi}{a^2}e^{-\frac{u^2}{2a^2}}e^{-\frac{v^2}{2a^2}}
    +\frac{\sqrt{2\pi}}{ap_v v^2}e^{-\frac{u^2}{2a^2}}
    +\frac{\sqrt{2\pi}}{ap_u u^2}e^{-\frac{v^2}{2a^2}}
    +\frac{2}{p_up_v u^2 v^2}}.\label{expoz}
\end{equation}
Each term was obtained by the residue theorem. The specific procedure for each term is detailed below:
\begin{itemize}
    \item First term : For the \(u\)- integration, we take the residue at the pole \(u_1=u_2\), followed by the pole at \(u_3=u_2\). For the \(v\)-integration, we take the residue at the pole \(v_1=v_2\), followed by the pole at \(v_3=v_2\).
    \item Second term : For the \(u\)- integration, we take the residue at the pole \(u_1=u_2\), followed by the pole at \(u_3=u_2\). For the \(v\)-integration, we take the residue at the pole \(v_1=v_2\), and then evaluate the integral over \(v_3\) using the saddle point approximation.
    \item Third term : For the \(u\)- integration, we take the residue at the pole \(u_1=u_3\), and then evaluate the integral over \(u_3\) using the saddle-point approximation. For the \(v\)-integration, we take the residue at the pole \(v_1=v_2\), followed by the pole at \(v_3=v_2\).
    \item Fourth term : For the \(u\)- integration, we take the residue at the pole \(u_1=u_3\), and then evaluate the integral over \(u_3\) using the saddle-point approximation. For the \(v\)-integration, we take the residue at the pole \(v_1=v_2\), and then evaluate the integral over \(v_3\) using the saddle point approximation.
\end{itemize}
The \(i\ep\) prescription plays a crucial role in the pole calculation, which indicates that the result depends on the operator ordering.
While we have considered wave packet states so far, we now construct a similar state that is obtained by exponentiating an operator. 
This state is (approximately) localized at $x=0$ at $t=0$.

Recall that the wave packet state was defined as: 
\begin{align}
    \ket{p,\omega} & = \int dt dx \, e^{-\frac{t^2+x^2}{2a^2}-i\omega t +ipx} \cO(t,x) \ket{0} \nn
    & = \Phi^\dagger\ket{0} ,
\end{align}
where we introduce the operator \(\Phi^\dagger\) as
\begin{align}
    \Phi^\dagger \coloneqq \int dtdx\, e^{-\frac{t^2+x^2}{2a^2}-i\omega t +ipx}\cO(t,x).
\end{align}
\(\Phi^\dagger\) acts like a creation operator, generating a wave packet with energy \(\omega\) and momentum \(p\), while \(\Phi\) acts like an annihilation operator. In fact, considering \(\bra{0}\Phi^\dagger \Phi\ket{0}\), we can easily see that it is suppressed by \(e^{-a^2(\omega^2+p^2)}\).
We construct a state \(\ket{U}\) 
as
\begin{align}
    \ket{U} \coloneqq e^{i \alpha \lr{\Phi+\Phi^\dagger} } \ket{0},
\end{align}
where \(\alpha\) is a small parameter, and
\begin{align}
    &\bra{U} \cO\ket{U} \nn
    =& \bra{0}\lr{1-i\alpha (\Phi+\Phi^\dagger)-\frac{\alpha^2}{2}\lr{\Phi+\Phi^\dagger}^2}\cO 
     \lr{1+i\alpha (\Phi+\Phi^\dagger)-\frac{\alpha^2}{2}\lr{\Phi+\Phi^\dagger}^2}\ket{0} \nn
    &+\cO(\alpha^3).
\end{align}
We note that if $\Phi$ is a local operator at $X$, then $\bra{U} \cO(Y) \ket{U}=\bra{0} \cO(Y) \ket{0}$ if $X$ and $Y$ are causally separated because of the microcausality, i.e., $[\Phi(X), \cO(Y)]=0$. 

Focusing on order \(\alpha^2\), 
\begin{align}
    &\bra{0}-\frac{\alpha^2}{2}\lr{\Phi+\Phi^\dagger}^2\cO -\frac{\alpha^2}{2}\cO\lr{\Phi+\Phi^\dagger}^2 +\alpha^2 \lr{\Phi+\Phi^\dagger}\cO\lr{\Phi+\Phi^\dagger}\ket{0}.
\end{align}
This term is nonzero only in calculations that reflect the order of \(\cO\) and \(\Phi\).%
\footnote{
Note that contributions from \(\bra{0}\cO (\Phi^\dagger)^2\ket{0}\) and \(\bra{0}\Phi^2\cO\ket{0}\) approximately vanish as following:
\begin{align}
    &\bra{0}\cO (\Phi^\dagger)^2\ket{0} \nn
    = & \int dx_1 dt_1 dx_3 dt_3 e^{-\frac{(x_1)^2+(t_1)^2}{2a^2}-i\omega t_1 +ipx_1}e^{-\frac{(x_3)^2+(t_3)^2}{2a^2}-i\omega t_3 + ipx_3} \langle 0| \cO(t,x)\cO(t_1,x_1) \cO(t_3,x_3)|0\rangle \nn
    = &\,\frac{1}{4} \int du_1 dv_1 du_3 dv_3 e^{-\frac{\lr{u_1+ia^2p_u}^2}{4a^2}}e^{-\frac{\lr{u_3+ia^2p_u}^2}{4a^2}}e^{-\frac{\lr{v_1+ia^2p_v}^2}{4a^2}}e^{-\frac{\lr{v_3+ia^2p_v}^2}{4a^2}}e^{-a^2p_u^2}e^{-a^2p_v^2} \nn
    & \times \frac{C_{123}}{\lr{u-u_1 -i \ep}^{\frac{\Delta}{2}}\lr{u_1-u_3 -i\ep}^{\frac{\Delta}{2}}\lr{u_3-u+i\ep}^{\frac{\Delta}{2}}\lr{v_1-v+i\ep}^{\frac{\Delta}{2}}\lr{v_3-v_1+i\ep}^{\frac{\Delta}{2}}\lr{v-v_3-i\ep}} \nn
    \simeq & \,0 ,
\end{align}
where we have used the fact that due to the change in the sign of \(\omega\) and the difference in ordering, the pole is no longer located inside the complex integration contour. 
This is immediately clear from the \(u_1\) integral.
}
However, the fourth term of \eqref{expoz} is calculated using the saddle-point approximation for \(u_2\) and \(v_2\), which does not reflect the \(i\ep\) prescription.%
\footnote{
\(u_2\) and \(v_2\) correspond to the insertion point of the operator \(\cO\).
}
In other words, it does not reflect the operator ordering of \(\cO\) and \(\Phi\). 
Therefore, while the fourth term in the calculation of a single matrix element $\langle0|\Phi\cO\Phi^\dagger|0\rangle$ gives a nonzero contribution at spacelike separation, this piece is absent for the state obtained by the unitary operator. 
For the state $\ket{U}$, the expectation value is a sum over all possible operator orderings. 
\MD{In this sum, the seemingly acausal fourth term is cancelled by analogous contributions from other orderings (such as $\langle0|\cO\Phi \Phi^\dagger|0\rangle$, etc.).}




\newpage


\providecommand{\href}[2]{#2}\begingroup\raggedright\endgroup

\end{document}